\documentclass[aps,prd,eqsecnum,showpacs,floatfix]{revtex4}
\usepackage{latexsym}
\usepackage{graphicx,subfigure}
\begin{document}

\thispagestyle{empty}

\title{Growth of primordial black holes in a universe 
containing a massless scalar field}
\author{Tomohiro Harada\footnote{Electronic
address:T.Harada@qmul.ac.uk}
and
B.~J.~Carr\footnote{Electronic
address:B.J.Carr@qmul.ac.uk}}
\affiliation{
Astronomy Unit, School of Mathematical Sciences, 
Queen Mary, University of London, 
Mile End Road, London E1 4NS, UK}
\date{\today}

\begin{abstract}                
The evolution of primordial black holes
in a flat Friedmann universe with 
a massless scalar field is investigated in fully general relativistic
numerical relativity.
A primordial black hole is expected to form with a scale comparable to the cosmological 
apparent horizon, in which case it may go through an initial 
phase with significant accretion.
However, if it is very close to the cosmological 
apparent horizon size, the accretion is suppressed due
to general relativistic effects.
In any case, it soon gets smaller than 
the cosmological horizon and 
thereafter it can be approximated
as an isolated vacuum solution with decaying mass accretion.
In this situation the dynamical and 
inhomogeneous scalar field is typically equivalent 
to a perfect fluid with a stiff equation of state $p=\rho$. 
The black hole mass never increases by more than a factor of two, despite recent claims that primordial black holes might grow substantially through accreting quintessence. It is found that the gravitational memory scenario,
proposed for primordial black holes 
in Brans-Dicke and scalar-tensor theories of 
gravity, is highly unphysical.
\end{abstract}
\pacs{04.70.Bw, 97.60.Lf, 04.25.Dm, 95.35.+d}
\maketitle
\section{Introduction}

The observed anisotropy and 
polarisation of the cosmic microwave background radiation
give very precise information about the cosmic history since
last scattering.
Using a simple physical
model, we can determine cosmological parameters
with very high accuracy and also learn about structure 
formation and reionisation.
Nevertheless, it is still difficult 
to obtain information about the universe 
before big bang nucleosynthesis
without making many extra assumptions.
In this context, primordial black holes (PBHs) could be 
one of the most important fossils of the very early universe.
Such black holes may have formed directly from
primordial density perturbations~\cite{hawking1971a} and may contribute to 
the cosmological $\gamma$-ray background radiation, the cosmic ray flux and
the dark matter density. This leads to important observational constraints
on the number of PBHs~\cite{carr1975} and hence on models of the very early universe.  

To understand these constraints,
it is important to know whether PBHs can accrete 
enough to become much more massive than they were at
formation. This topic has a long history. It was originally claimed, on the basis of a simple Newtonian 
argument, that a PBH much smaller than the cosmological particle horizon at formation would not
accrete much at all, whereas one comparable to the horizon size would continue to grow at the same rate as the horizon until the end of the radiation-dominated era~\cite{zn1967}. Since one would {\it expect} a PBH to be of order the particle horizon size at formation in order to collapse against the pressure, this suggests that  all PBHs might grow to the horizon mass at the end of the radiation era, which is about $10^{15}$ solar masses.  Since there is no evidence for such enormous black holes in the universe today, the conclusion seemed to be that PBHs never formed.

However, this conclusion was very suspect. If PBHs really could grow as fast as the particle horizon, there would have to exist spherically symmetric self-similar solutions to Einstein's equations containing black holes in an exact Friedmann background.
A study of solutions of this kind containing radiation (i.e. fluid with equation of state $p=\rho /3$) showed that  there are no self-similar PBH
solutions if the black hole is formed by purely local processes (i.e. if
the background is exactly Friedmann beyond some radius). Self-similar
solutions are possible only if the initial perturbation of the 
Friedmann background extends to infinity~\cite{ch1974} . Since the PBH must 
soon become much smaller than the
cosmological horizon, at which point the Newtonian argument {\it should} be applicable,  this suggests that it will only increase its mass by a small amount. 

This conclusion was subsequently extended to more general fluids, with an equation of state of the form $p=k \rho$~\cite{carr1976,bh1978a}. Although there was an initial claim that self-similar growth might be possible in the special case of a stiff fluid ($p=\rho$)~\cite{lcf1976}, this claim was subsequently challenged~\cite{bh1978b}. It was found to be true only in rather contrived circumstances in which the stiff fluid is converted into radiation at the black hole's event horizon.  
Therefore the conclusion that there are no self-similar 
solutions containing PBHs formed by purely local processes seems to be true very generally. See~\cite{cc2000} for a review of self-similar solutions. This conclusion is also confirmed by numerical calculations. If the formation and evolution of a PBH in a general fluid universe 
with a local perturbation is simulated numerically, without assuming 
self-similarity, it is found that the PBH 
soon becomes much smaller than the
cosmological horizon and this excludes self-similar 
growth~\cite{nnp1978,np1980}. More recently, PBH 
formation from density perturbations 
has been investigated in the context of 
critical phenomena~\cite{nj1999,jn1999,ss1999,hs2002,mmr} but again with no evidence for self-similar growth.

Hitherto studies of PBHs have mainly focused on perfect fluid universes with equation of state $p=k\rho$.
However, it is also natural to consider a universe whose density is dominated by a scalar field.
For example, in the chaotic inflation scenario it is postulated that there is a 
pre-inflationary stage in which
the scalar field moves randomly in space and time.
In the pre-heating inflationary scenario, 
it is also natural to consider a scalar-field-dominated 
era~\cite{kls1994}. 
More recently, the study of PBHs in the quintessence scenario 
has attracted attention~\cite{frol2004}.  
It is therefore important to examine whether the conclusion that PBH accretion is small also applies in this case. 

If the scalar field is massless and there is no scalar potential, then it is well known~\cite{madsen1988}  that it  is equivalent to 
a stiff fluid providing the gradient of the scalar field remains timelike, as usually applies. The fact that  there is no self-similar solution in the stiff case therefore suggests that accretion is also limited in the scalar field case. Despite this, the original Newtonian argument 
for self-similar growth has recently been applied in the quintessence scenario to argue that PBHs could grow enough to form the supermassive black holes found in galactic nuclei~\cite{bm2003}. A generalization of this analysis~\cite{ch2005}, not necessarily involving self-similarity but still based on the Newtonian analysis, has also claimed there could be appreciable quintessence accretion in some circumstances. Both these analyses contravene the conclusions of the earlier work discussed above. However, since the earlier work did not strictly include the scalar field case and did not allow for a scalar potential, a more careful analysis is required before concluding that these analyses are erroneous. 

Scalar fields are also relevant in Brans-Dicke and scalar-tensor theories of 
gravity, where the gravitational ``constant'' varies in 
space and time. This is because, if
there is a single gravitational scalar field, such theories  
can be transformed into the usual
Einstein gravity with a single scalar field~\cite{de1992}.
Brans-Dicke and scalar-tensor theories are particularly relevant to PBHs since the black holes may form when $G$ was very different from today. Indeed the ``gravitational memory'' scenario has been proposed~\cite{barrow1992},
in which the value of $G$ within the black hole
is assumed to be preserved as the cosmological 
background value evolves. The observational constraints on PBHs 
depend strongly on whether or not one has gravitational memory~\cite{bc1996}.
It is not clear whether this applies but, if it does, it should be due
to the properties of the black hole event horizon rather than 
those of the matter fields involved. 

It is interesting that the problem of gravitational memory is also closely related to the problem of accretion~\cite{cg1999}. This is because it turns out that appreciable accretion of the scalar field energy is required for gravitational memory to be preserved and so the issue of whether there is a self-similar solution again becomes relevant. Indeed Carr and Goymer argued that there is unlikely to be gravitational memory on the grounds that there is no self-similar solution in the stiff fluid case~\cite{cg1999}. 
Evidence against the gravitational memory scenario, at least in particular situations, has also been obtained using other arguments~\cite{jacobson1999,hgc2002}. 

As a first step to investigating these questions numerically, in this paper we consider a universe containing a massless scalar field but no matter. To implement the simulations,
we use a double-null formulation
of the Einstein equations based on the work by 
Hamad\'e and Stewart~\cite{hs1996}. This has been shown to be a very powerful tool for investigating 
critical collapse~\cite{hs1996} and 
the internal structure of black holes~\cite{sp1999}. 
This study represents an improvement on earlier work~\cite{hgc2002}, in which the gravitational effect of the scalar field was neglected, since it uses the full field equations. It should also be seen in conjunction with two other papers~\cite{hc2004a,hc2004b}; we consider general constraints on the size of a PBH at formation in the first 
and some effects associated with very large PBHs in the second.

The plan of the paper is as follows. In Section II, we describe the double-null formulation and present the basic equations for the scalar field case. In Section III, we discuss how we set up the initial data for the simulations. Section IV presents our results, with particular emphasis on the evolution of the black hole event horizon and the spatial profile of the scalar field.  In Section V, we discuss the implications of these results for the accretion and gravitational memory issues, also considering the applicability of the Newtonian accretion formula.  In Section VI, we draw some general conclusions. Details of the numerical code and particular exact solutions are described in Appendices. We adopt units in which $G=c=1$ and the abstract index notation of reference~\cite{wald1983}.

\section{Double-null formulation of the Einstein equations}
\label{sub:double-null}
We consider a massless scalar field in 
general relativity, for which 
the stress-energy tensor is
\begin{equation}
T_{ab}=\Psi_{,a}\Psi_{,b}-\frac{1}{2}g_{ab}\Psi^{,c}\Psi_{,c}.
\end{equation}
The Einstein equations are
\begin{equation}
R_{ab}-\frac{1}{2}g_{ab}R=8\pi T_{ab},
\label{eq:einstein}
\end{equation}
and the equation of motion for the scalar field is
\begin{equation}
\Box\Psi=\Psi^{;a}_{~~;a}=0.
\label{eq:eom}
\end{equation}
We also focus on a spherically symmetric system,
for which the line element can be written in the form
\begin{equation}
ds^{2}=-a^{2}(u,v)dudv+r^{2}(u,v)(d\theta^{2}+\sin^{2}\theta d\phi^{2}),
\end{equation}
where $u$ and $v$ are advanced and retarded time coordinates,
respectively, and $r$ is the ``area radius'' (the proper area of the sphere
of constant $r$ being $4\pi r^2$).
Eqs.~(\ref{eq:einstein}) and (\ref{eq:eom}) then imply that
we have 14 first-order partial differential equations
and two auxiliary equations (see Section 2 of~\cite{hs1996}).
By adopting a double-null coordinate choice, we can 
simulate regions outside the cosmological apparent horizon and 
inside the black hole apparent horizon 
simultaneously.

In spherically symmetric spacetimes the existence and position of apparent horizons can be inferred from the form of the Hawking mass. This is a well-behaved quasi-local mass, which can be written as
\begin{equation}
m=\frac{r}{2}\left(1+\frac{4r_{,u}r_{,v}}{a^{2}}\right).
\label{eq:misner_sharp}
\end{equation}
Using the above equation
and Eq. (2.6) of~\cite{hs1996},
we can derive the following useful relations:
\begin{eqnarray}
m_{,u}&=&-\frac{8\pi r^{2}r_{,v}(\Psi_{,u})^{2}}{a^{2}} ,
\label{eq:m_u}\\
m_{,v}&=&-\frac{8\pi r^{2} r_{,u}(\Psi_{,v})^{2}}{a^{2}}.
\label{eq:m_v}
\end{eqnarray}
A region is trapped if $r_{,u}<0$ and $r_{,v}<0$,
while it is antitrapped if $r_{,u}>0$ and $r_{,v}>0$.
The black hole and cosmological apparent horizons
are defined as marginally trapped and anti-trapped surfaces,
respectively.
Providing the black hole horizon is within the cosmological horizon, 
these correspond to the conditions 
$r_{,v}=0$ and $r_{,u}=0$, respectively. 
Thus the relation $r=2m$ is satisfied on both apparent horizons because $r_{,u}r_{,v}=0$ there.
We confine attention to this situation in the present paper. However, as discussed in a separate paper~\cite{hc2004b}, in some circumstances the black hole horizon can be outside the cosmological horizon and the situation is then more complicated.
In this case, we can still use two different marginal surfaces
on which $r_{,v}=0$ and $r_{,u}=0$, respectively~\cite{hayward1996}. 
However, these are
no longer everywhere identified with the black hole and cosmological apparent horizons.

Note that the black hole and cosmological apparent horizons are
distinct from the black hole event horizon and cosmological particle
horizon, which are always null. For numerical purposes, the apparent horizons are 
easier to find than the event and 
particle horizons because the complete spacetime, including the initial 
singularity, is needed to identify the latter,
whereas the former can be found from the quasi-local properties 
of spacetime alone.
However, we will also discuss the behaviour of the black hole event horizon
in Section~\ref{sec:results}.

\begin{figure}[htbp]
\includegraphics[scale=1]{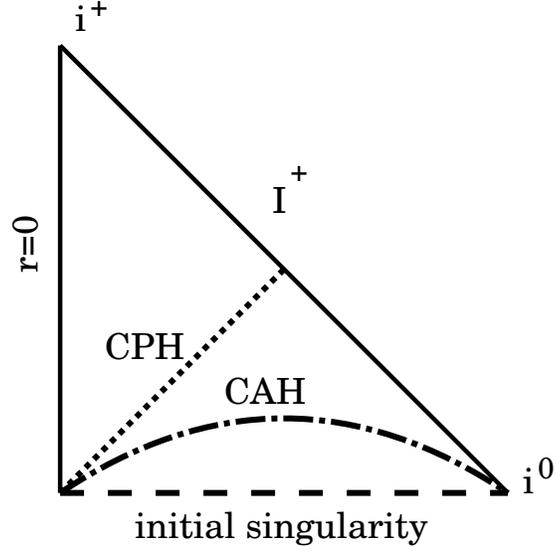}
\caption{\label{fig:flat_friedmann}
The conformal diagram of the flat Friedmann 
spacetime with a massless scalar field.
The cosmological apparent horizon is spacelike and outside 
the cosmological particle horizon.}
\end{figure}

\begin{figure}[htbp]
\includegraphics[scale=1]{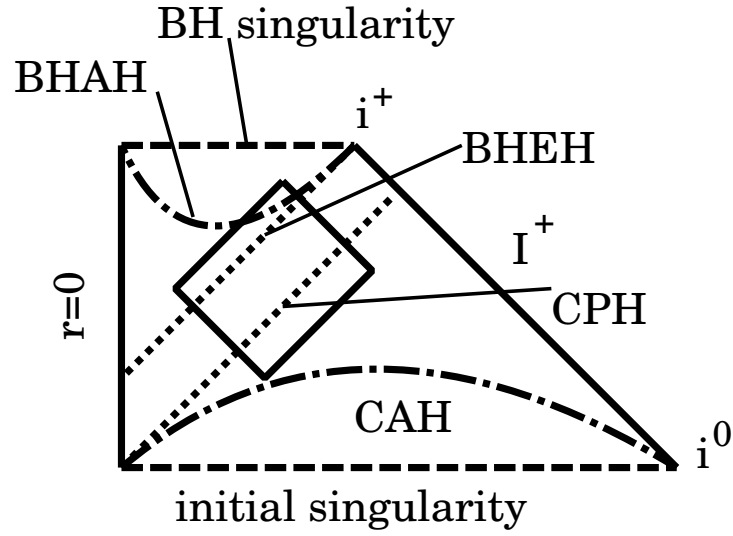}
\caption{\label{fig:pbh}
The conformal diagram of the possible causal structure of the calculated model.
The region enclosed by a diamond is the calculated region 
$[u_{0},u_{1}]\times [v_{0},v_{1}]$.}
\end{figure}

We first consider the flat Friedmann model, for which $a^{2}(u,v)$, $r(u,v)$
and $m(u,v)$ are given explicitly 
in terms of the standard double null coordinates
in Appendix~\ref{sec:exact_solutions}. The condition $r=2m$ implies that the cosmological  apparent horizon
is $3u+v=0$ and this implies that its radius is just the Hubble scale
$H^{-1}$ in a flat Friedmann universe, as described in~\cite{hc2004a}. On the other hand, the cosmological particle horizon is clearly $u=0$, corresponding to a photon propagating outwards from the big bang. This implies that the apparent horizon is always outside the particle horizon and the conformal diagram of the spacetime is as indicated in Fig.~\ref{fig:flat_friedmann}. This also shows the initial (big bang) spacelike singularity. If one perturbs such a spacetime but without introducing a black hole, the conformal diagram remains the same but the trajectories of constant space and time coordinate change. If one has a black hole embedded in an exact or asymptotically flat Friedmann model, the conformal diagram will change to the form indicated in Fig.~\ref{fig:pbh}. There are now also a black hole event horizon and apparent horizon, although one cannot give explicit analytic expressions for these, and a final (black hole) spacelike singularity.

 If we define time and space coordinates, $T$ and $X$, by
\begin{eqnarray}
u&=&T-X,\\ 
v&=&T+X,
\end{eqnarray}
we can put the 2-dimensional part of the metric tensor 
into conformally flat form. 
As in the usual 3+1 approach,
we define the energy density $\rho_{\rm H}$ and 
momentum density $J$ measured by an observer moving normal 
to the $T=\mbox{const}$ spacelike 
hypersurface by
\begin{eqnarray}
\rho_{\rm H}&\equiv& T_{ab}\hat{T}^{a}\hat{T}^{b}, \\
J&\equiv& -T_{ab}\hat{T}^{a}\hat{X}^{b}, 
\end{eqnarray}
where $\hat{T}^{a}$ and $\hat{X}^{b}$ are unit vectors
parallel to $T^{a}$ and $X^{b}$, respectively.

It should be stressed that the above description is observer-dependent and one can
adopt another point of view.
Providing the scalar field is vorticity-free and 
has a timelike gradient,
it is equivalent to a perfect fluid with a ``stiff''
equation of state $p=\rho$~\cite{madsen1988}.
The stress-energy tensor is then 
\begin{equation}
T^{ab}=\rho_{\rm S}(g^{ab}+2U^{a}U^{b}),
\end{equation}
where the energy density $\rho_{\rm S}$ and the 4-velocity $U^{a}$
of the stiff fluid are
\begin{eqnarray}
\rho_{\rm S}&=&-\frac{1}{2}\Psi_{,a}\Psi^{,a}, \\
U_{a}&=&\frac{-\Psi_{,a}}{\sqrt{-\Psi^{,b}\Psi_{,b}}}.
\end{eqnarray}
It should be noted that the 4-velocity of the 
stiff fluid is parallel to the (timelike) gradient 
of the scalar field.
We can define an observer-independent velocity function by
\begin{equation}
V\equiv \frac{dr}{d\tau}=r_{,a}U^{a},
\end{equation}
this being the rate of increase of $r$ per unit proper time 
$\tau$ along the worldline of the equivalent stiff fluid element.
Appendix~\ref{sec:basic_equations} gives expressions 
for these physical quantities
in terms of the quantities calculated numerically.

\section{Initial data for PBHs}
\subsection{Structure of initial data}
\begin{figure}[htbp]
\includegraphics[scale=1]{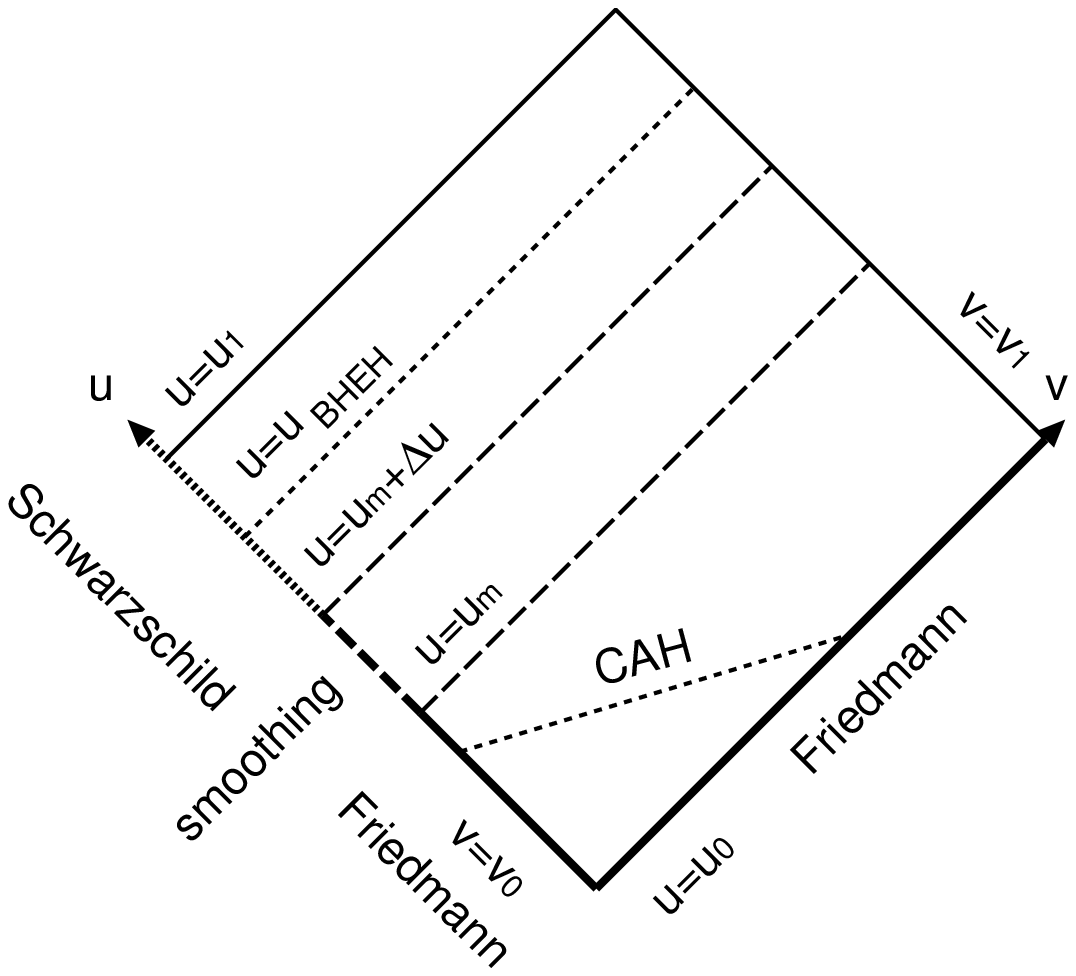}
\caption{\label{fig:local_perturbation}
Schematics figure of the initial data setting.
The relation between $u_{m}$ and CAH is 
true only for Models A-C.}
\end{figure}

The initial data are prescribed on the outgoing null surface $u=u_{0}$
and the ingoing null surface $v=v_{0}$.
The region of calculation is the diamond 
$[u_{0},u_{1}]\times [v_{0},v_{1}]$ shown 
in Fig.~\ref{fig:local_perturbation}, which is also equivalent to the diamond in Fig.~\ref{fig:pbh}.
We have three independent functions
on the two null surfaces: $a^{2}$, $\Psi$ and $r$. Two of them can be chosen freely
and the other one is determined by the initial value
equations on the null surfaces.
It is convenient to choose
\begin{equation}
a^{2}(u_{0},v),a^{2}(u,v_{0}),\Psi(u_{0},v),\Psi(u,v_{0})
\end{equation}
as the free initial data and to regard
\begin{equation}
r(u_{0},v),r(u,v_{0})
\end{equation}
as being determined
by the initial value equations.
We can regard $\Psi(u_{0},v)$ and $\Psi(u,v_{0})$ 
as the physical degrees of freedom in the initial data,
while the choice for $a^{2}(u_{0},v)$
and $a^{2}(u,v_{0})$ fixes the gauge.

It is usually assumed that the perturbation from which the PBH forms is local, so that the universe is exactly Friedmann at sufficiently large distances. 
Since information propagates at the speed of light,
the matching between the perturbed and
flat Friedmann regions always corresponds to some outgoing null ray
$u=u_{\rm m}=\mbox{const}$ and such a ray is shown in Fig.~\ref{fig:local_perturbation}.
This means that the region outside the matching surface $u=u_{\rm m}$
is always described by the flat Friedmann solution.
On the outgoing null surface $u=u_{0}$, we therefore assume that
the initial data are given by 
the flat Friedmann solution for $v_{0}\le v\le v_{1}$.
On the ingoing null surface $v=v_{0}$, we assume they are
given by the flat Friedmann solution 
for $u_{0}\le u< u_{\rm m}$
but by some perturbation of it for $u_{\rm m}\le u\le u_{1}$,
where an appropriate matching is implemented at $u=u_{\rm m}$. As described in the next section, the perturbed region contains a Schwarzschild black hole. However, since this is a vacuum solution, it only applies instantaneously at $v=v_{0}$ because the inflowing matter will fill the vacuum immediately.

If the perturbation is assumed to be local, 
we can derive various upper limits on the black hole size~\cite{hc2004a}. These are generally within the cosmological particle horizon, which would then correspond to the line $u=0$ in Fig.~\ref{fig:local_perturbation}.
However, a perturbation can be larger than the 
cosmological particle horizon in some circumstances -- indeed most of the cases we consider in this paper have this feature. The particle horizon is then within the perturbed region and no longer given by $u=0$. In some circumstances, e.g. for
quantum fluctuations resulting from inflation, the perturbation may extend indefinitely, or at least well beyond the usual particle horizon.  Generally the only upper limit on the size of the perturbation 
comes from the requirement that the the perturbed region must be
part of our universe rather a separate closed universe. 
The condition for this has been derived precisely for 
the situation in which the collapsing region is homogeneous 
and the equation of state is $p=k\rho$~\cite{hc2004a}.  In this case, the closed-universe scale is always well beyond the cosmological apparent horizon size. 
Even if the perturbed region is within our universe, the solution may exhibit anomalous features if it is too large. These anomalies are described elsewhere~\cite{hc2004b} but we do not choose initial data which  lead them to arise in this paper.

\subsection{Setting up the initial data}
In the flat Friedmann region, 
we choose the coordinate system 
given in Appendix~\ref{subsec:flat_friedmann}.
We also impose flat Friedmann initial data for $a^{2}$ and $\Psi$
on the initial outgoing null surface $u=u_{0}$ with $v_{0}\le v\le v_{1}$.
This is also given in Appendix~\ref{subsec:flat_friedmann}.
On the initial ingoing null surface $v=v_{0}$, we choose flat Friedmann data for $a^{2}$ for $u_{0}\le u \le u_{1}$.
For the function $\Psi$, 
we use the same data 
on the initial ingoing null surface
for $u_{0}\le u\le u_{\rm m}$, 
but $\Psi=\mbox{const}$ for $u_{\rm m}+\Delta u<u\le u_{1}$.
This is equivalent to Schwarzschild data
in coordinates penetrating the black hole 
(cf. Appendix~\ref{subsec:schwarzschild}).
The sudden transition from flat Friedmann data to 
Schwarzschild data results in a discontinuity 
at $u=u_{\rm m}$, which reduces the numerical 
accuracy. Hence we smooth the transition with some
smoothing length $\Delta u$; we use a quadratic function 
between $u_{\rm m}$ and $u_{\rm m}+\Delta u$,
so that $\Psi$ and $\Psi_{,u}$ are continuous.
More precisely, we impose the following initial data 
for $a^{2}$ and $s\equiv \sqrt{4\pi}\Psi$:
\begin{eqnarray}
a^{2}(u,v_{0})&=&C^{2}\left(\frac{u+v_{0}}{2}\right) 
\label{eq:A_on_v0},\\
s(u,v_{0}) 
&=&
\left\{
\begin{array}{ll}
\displaystyle\frac{\sqrt{3}}{2}\ln\left(\displaystyle\frac{u+v_{0}}{2}\right)+s_{0}& \quad (u<u_{\rm m})\\
\displaystyle\frac{\sqrt{3}}{2}\left[\displaystyle\frac{(\Delta u)^{2}-(u_{\rm m}+\Delta u-u)^{2}}
{2\Delta u (u_{\rm m}+\Delta u)}+\ln\left(
\displaystyle\frac{u_{\rm m}+v_{0}}{2}\right)\right]+s_{0}&
\quad 
(u_{\rm m}\le u<u_{\rm m}+\Delta u) \\
\displaystyle\frac{\sqrt{3}}{2}\left[\displaystyle\frac{\Delta u}
{2(u_{\rm m}+v_{0})}+
\ln\left(\displaystyle\frac{u_{\rm m}+v_{0}}{2}\right)\right]+s_{0}&
\quad (u\ge u_{\rm m}+\Delta u)
\end{array}
\right. ,
\label{eq:s_on_v0}
\end{eqnarray}
on the initial ingoing null surface $v=v_{0}$,
and 
\begin{eqnarray}
a^{2}(u_{0},v)&=&C^{2}\left(\frac{u_{0}+v}{2}\right), \\
s(u_{0},v)&=&\frac{\sqrt{3}}{2}\ln\left(\frac{u_{0}+v}{2}\right)+s_{0},
\end{eqnarray}
on the initial outgoing null surface $u=u_{0}$.
Here $C$ and $s_0$ are constants and, without loss of generality, 
we can choose $C=1$ and $s_{0}=0$.
Note that one has a Schwarzschild vacuum for $u \ge u_{\rm m} + \Delta u$. However, as noted above, this situation only applies instantaneously at $v=v_0$, since the vacuum will be filled with the scalar field at any later or earlier time. Nevertheless, one expects the event horizon to evolve smoothly, so it will still be described by a single null line, as illustrated qualitatively in Fig.~\ref{fig:local_perturbation} 

To summarise, two parameters describe the initial conditions:
the location of the matching surface and the 
smoothing length. 
The region outside the matching surface is the exact 
flat Friedmann solution, so the first parameter
determines the mass of the perturbed region.
The second parameter determines the mass of the 
black hole compared to the perturbed mass.
A small smoothing length means a narrow boundary
between the exterior Friedmann region and 
the interior vacuum region. In this case, the mass inside
the vacuum region is nearly the same as 
that within the matching surface and so a large mass is 
concentrated at the central singularity.
A large smoothing length means a wide boundary,
so the mass inside the vacuum region 
is much smaller than this and so
a small mass is concentrated at the central singularity.
For vanishing smoothing length,
the black hole mass would be the same as 
that at the matching surface, so $u_m$ would also be the black hole event horizon, although it is difficult to implement this numerically because of the strong discontinuity.

\section{Results}
\label{sec:results}
The numerical code is based on Hamad\'e and Stewart~\cite{hs1996},
although it is slightly modified so as to be 
accurate to second-order (see
Appendix~\ref{sec:basic_equations}). 
As indicated in 
Fig.~\ref{fig:local_perturbation},
the calculated region is the diamond contained 
by $u_{0}$, $v_{0}$, $u_{1}$ and 
$v_{1}$.
The initial data are prescribed on the ingoing null surface $v_{0}=1$
and the outgoing null surface $u_{0}=-0.5$.
The values $u_{0}$ and $v_{0}$ fix the units.
For example, the background value for 
the Hubble parameter at $(u_{0},v_{0})$ is 4
and the Hubble length becomes 0.25 from Eq.~(\ref{eq:hubble}).
The value of $u_1$ is always taken to be 1.1 but $v_1$ varies between 2 and 4. 
On the initial ingoing null surface,
we make the matching at three values of $u_{\rm m}$ in the range $-0.5$ to 0.
We also use small and large smoothing lengths, corresponding to $\Delta u=0.02$ and $\Delta u=0.5$, respectively.

As shown in Appendix~\ref{subsec:flat_friedmann}, the cosmological apparent and particle horizons are given by $u_{\rm CAH}=-v/3$ and $u_{\rm CPH}=0$, respectively, in the flat Friedmann model.
As time proceeds, the outgoing null rays will become ever
more sensitive to $r$ near the black hole event horizon,
so the calculation 
is stopped at the value of $v$ at which the outgoing null rays near the 
black hole event horizon become too coarse to resolve. This then specifies the value $v_{1}$ in Fig.~\ref{fig:local_perturbation}. 
As shown in Appendix~\ref{subsec:flat_friedmann}, the radius of the unperturbed cosmological 
apparent horizon and the mass within it are given by
\begin{equation}
m_{\rm CAH}=r_{\rm CAH}/2=C v^{3/2}/(3\sqrt{3}).
\end{equation}
However, this equation must be modified if the perturbed region extends beyond the usual cosmological apparent horizon and this applies in one of the cases considered. 

The model parameters and initial black hole to cosmological 
horizon mass ratios
are summarised in Table~\ref{tb:models}.
We present results for four models.
For Model A, a sharp matching is made at the 
cosmological apparent horizon.
For Model B, a sharp matching is made
at the cosmological particle horizon,
which is well inside the cosmological apparent horizon. 
For Model C, a smooth matching is made 
at the cosmological apparent horizon.
For Model D, a smooth matching is made 
well outside the cosmological apparent horizon, although the black hole
itself is inside it. 
In a separate paper~\cite{hc2004b}, we consider
models in which a sharp matching is made well outside the cosmological
apparent horizon.
These models have very
different qualitative features, even the conformal diagram being 
modified, which is why we consider them separately. 

\subsection{Black hole horizons}
\begin{figure*}[htbp]
\begin{tabular}{cc}
\subfigure[]{\includegraphics[scale=0.45]{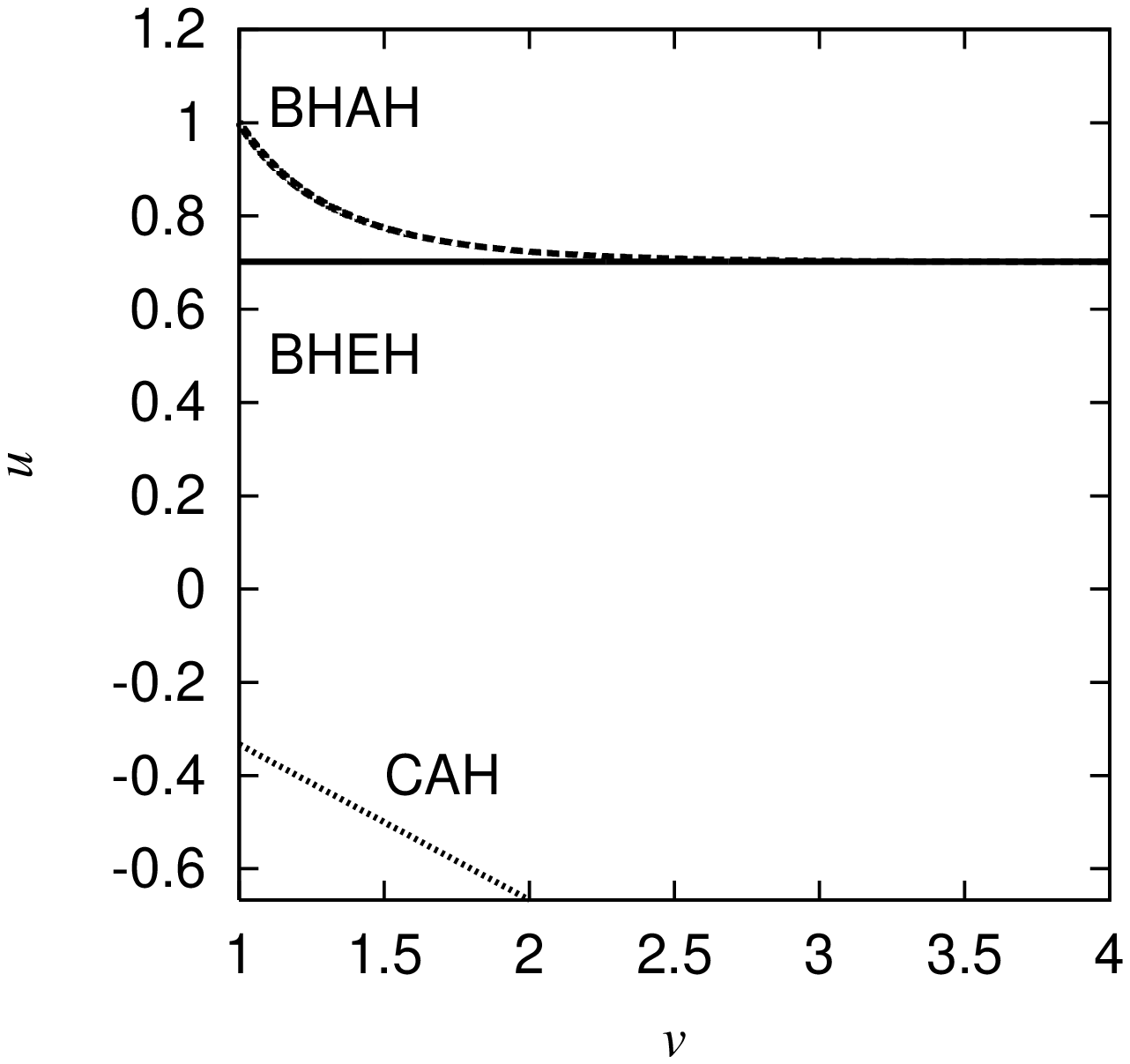}}&
\subfigure[]{\includegraphics[scale=0.45]{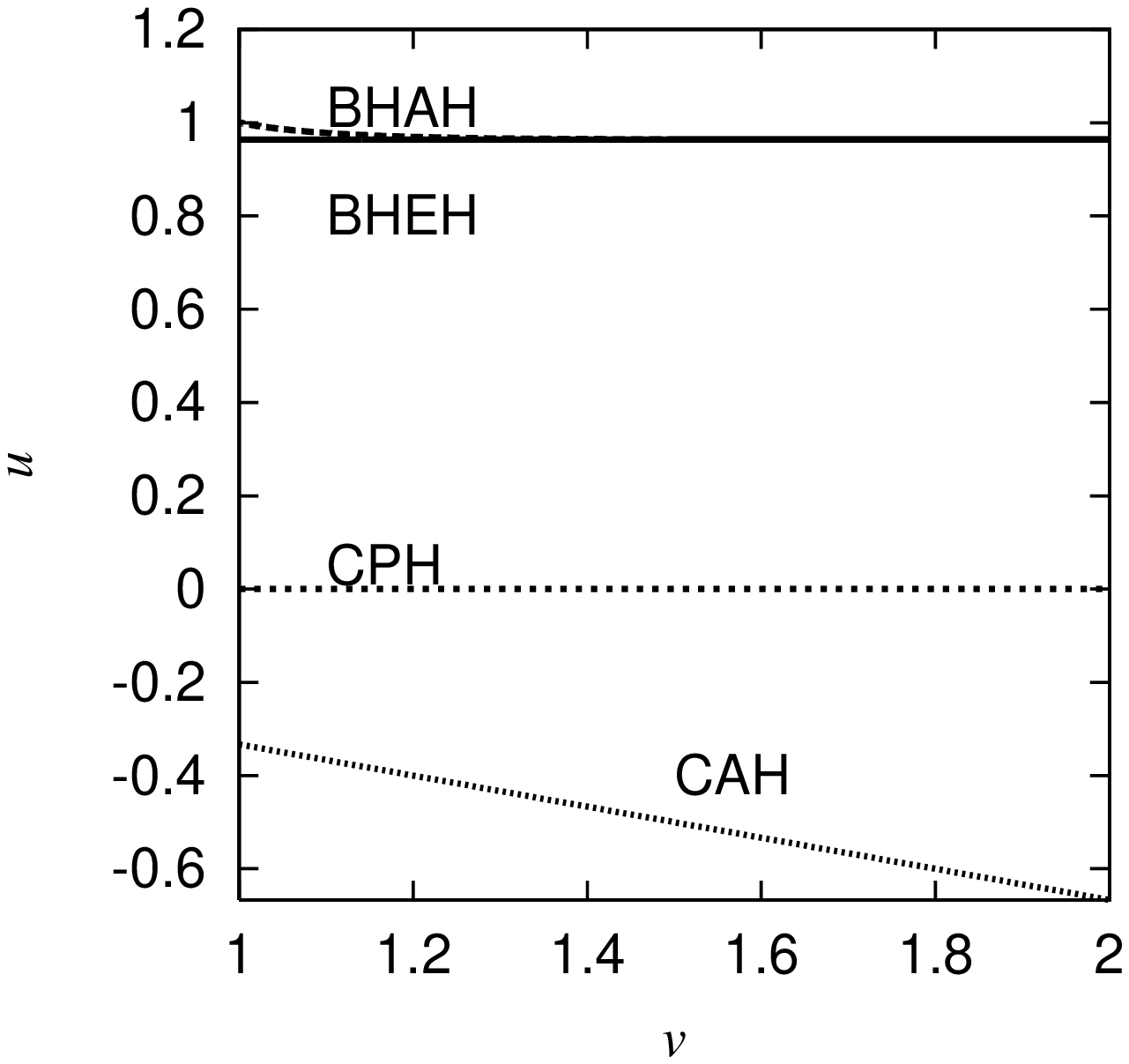}}\\
\subfigure[]{\includegraphics[scale=0.45]{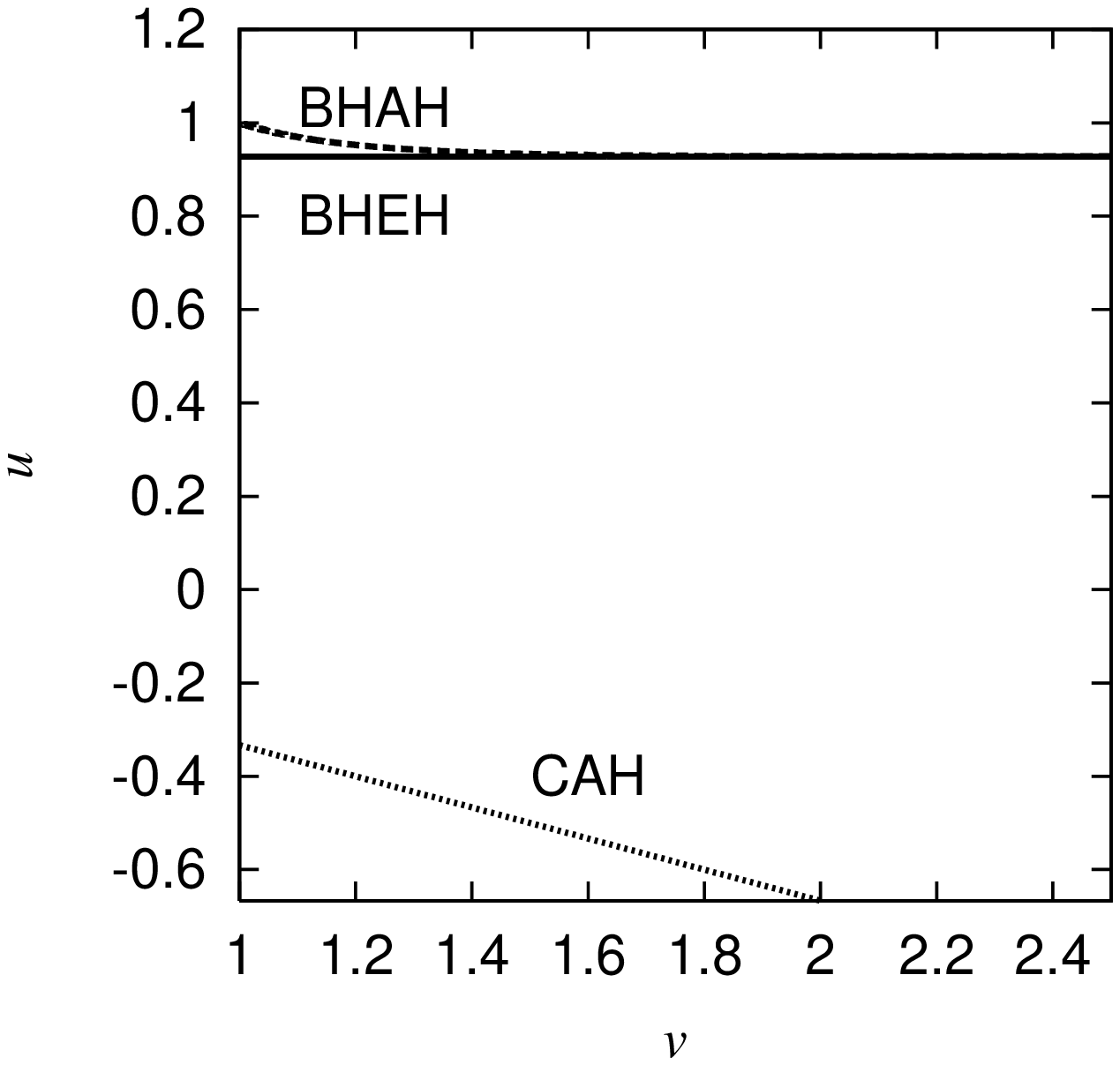}}&
\subfigure[]{\includegraphics[scale=0.45]{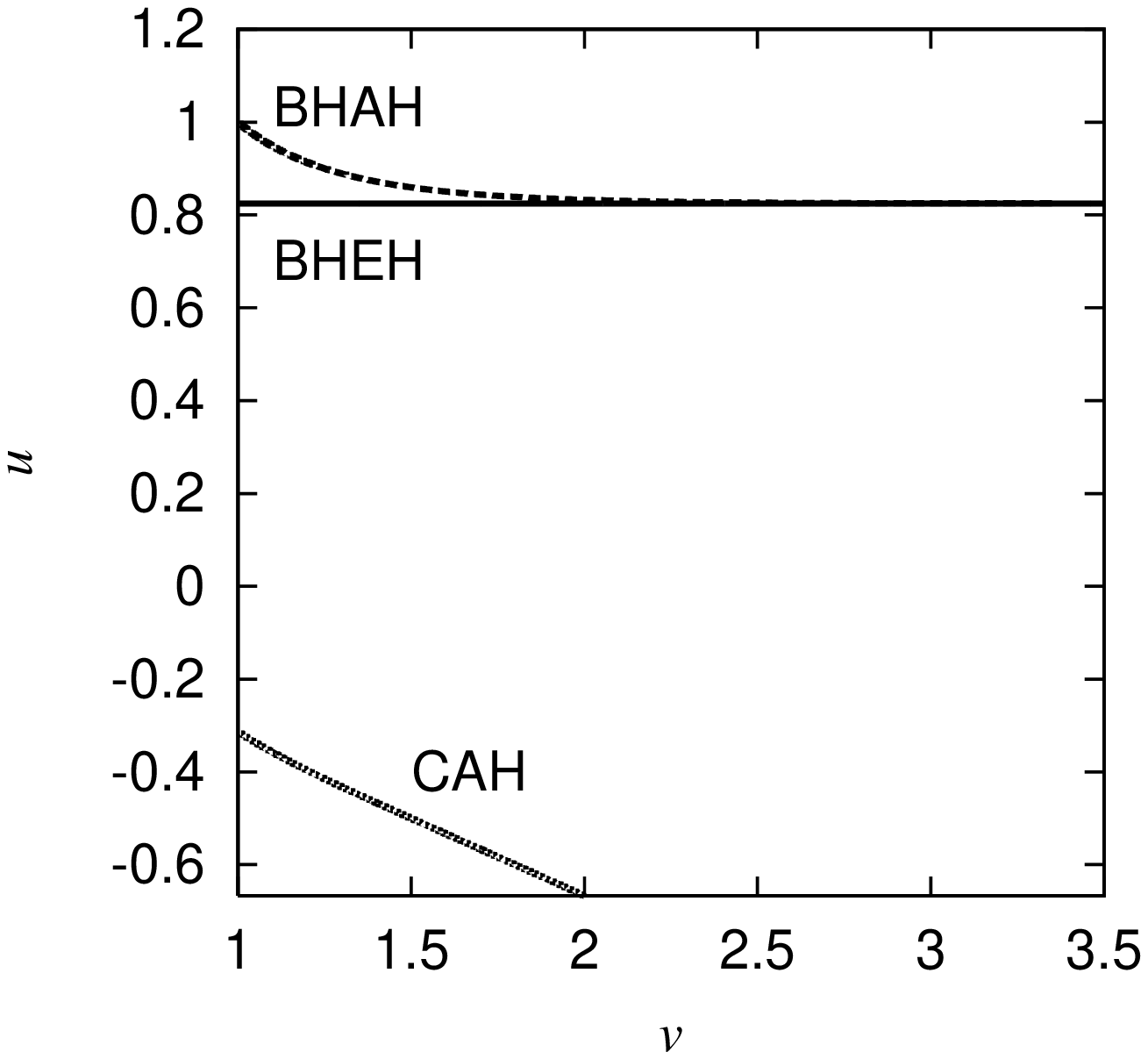}}\\
\end{tabular}
\caption{\label{fig:horizons_uv}
The locations of the black hole event horizon, black hole apparent horizon,
and cosmological apparent horizon in the $(u,v)$ plane for Models A--D
are plotted in (a)--(d), respectively.
For Model B, the location of the cosmological particle horizon
of the unperturbed flat Friedmann solution is also plotted.}
\end{figure*}

Figure~\ref{fig:horizons_uv} shows the locations 
of the black hole event horizon and apparent horizon
and the cosmological apparent horizon
in the $(u,v)$ plane for the four models.
Each figure can be identified with the diamond in Fig.~\ref{fig:local_perturbation}, tilted through 45 degrees. In all cases, 
it is seen that the black hole apparent horizon is initially well within the event horizon
but approaches it as time proceeds. 
In Fig.~\ref{fig:horizons_uv}(b), the location of the (unperturbed) 
cosmological particle horizon of the exact flat Friedmann 
solution is also plotted.
In other cases, the cosmological particle horizon is perturbed, so 
we do not plot it. The cosmological apparent horizon is unperturbed except in case D. 

\begin{figure}[htbp]
\begin{tabular}{cc}
\subfigure[]{\includegraphics[scale=0.45]{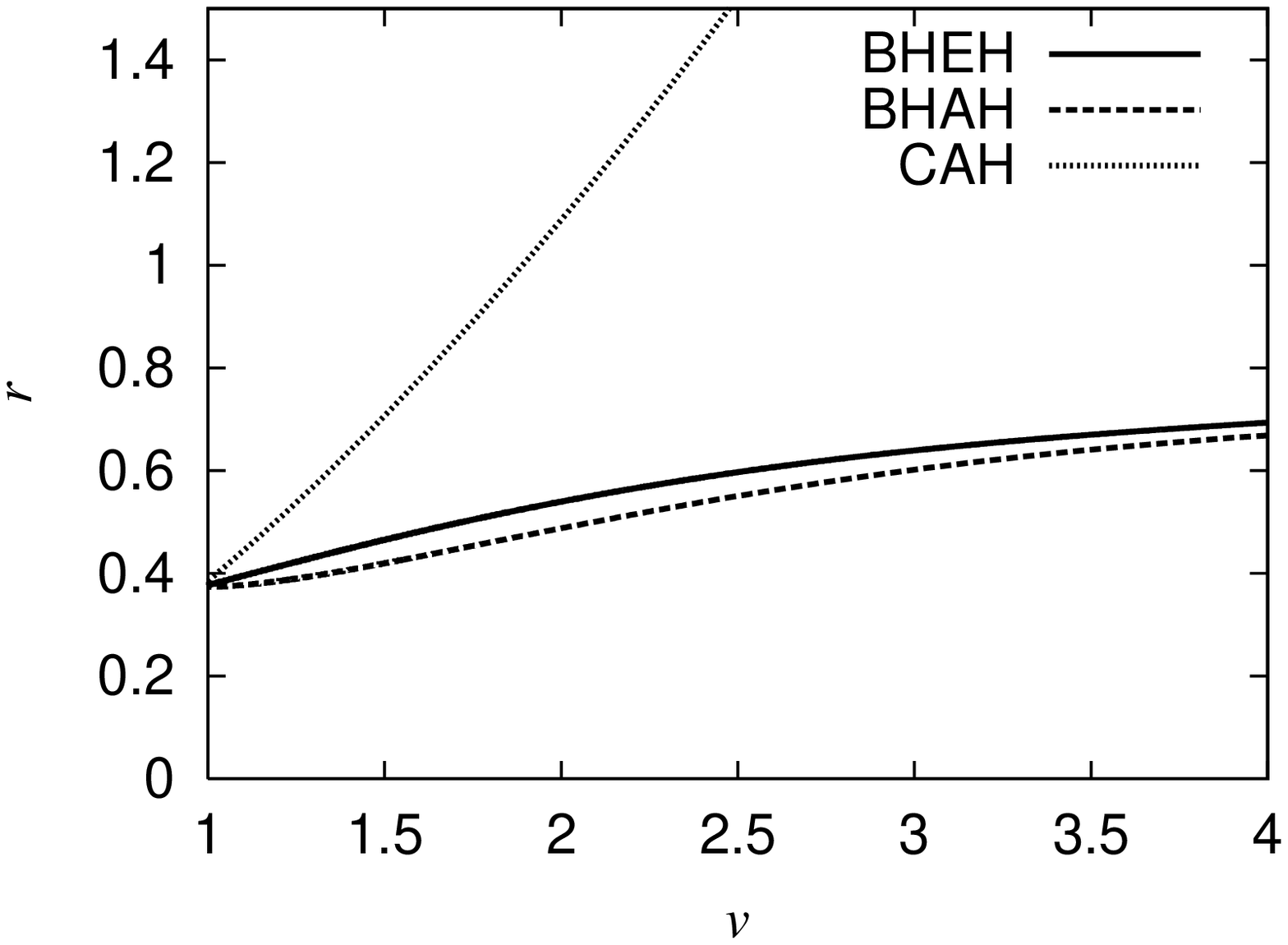}}&
\subfigure[]{\includegraphics[scale=0.45]{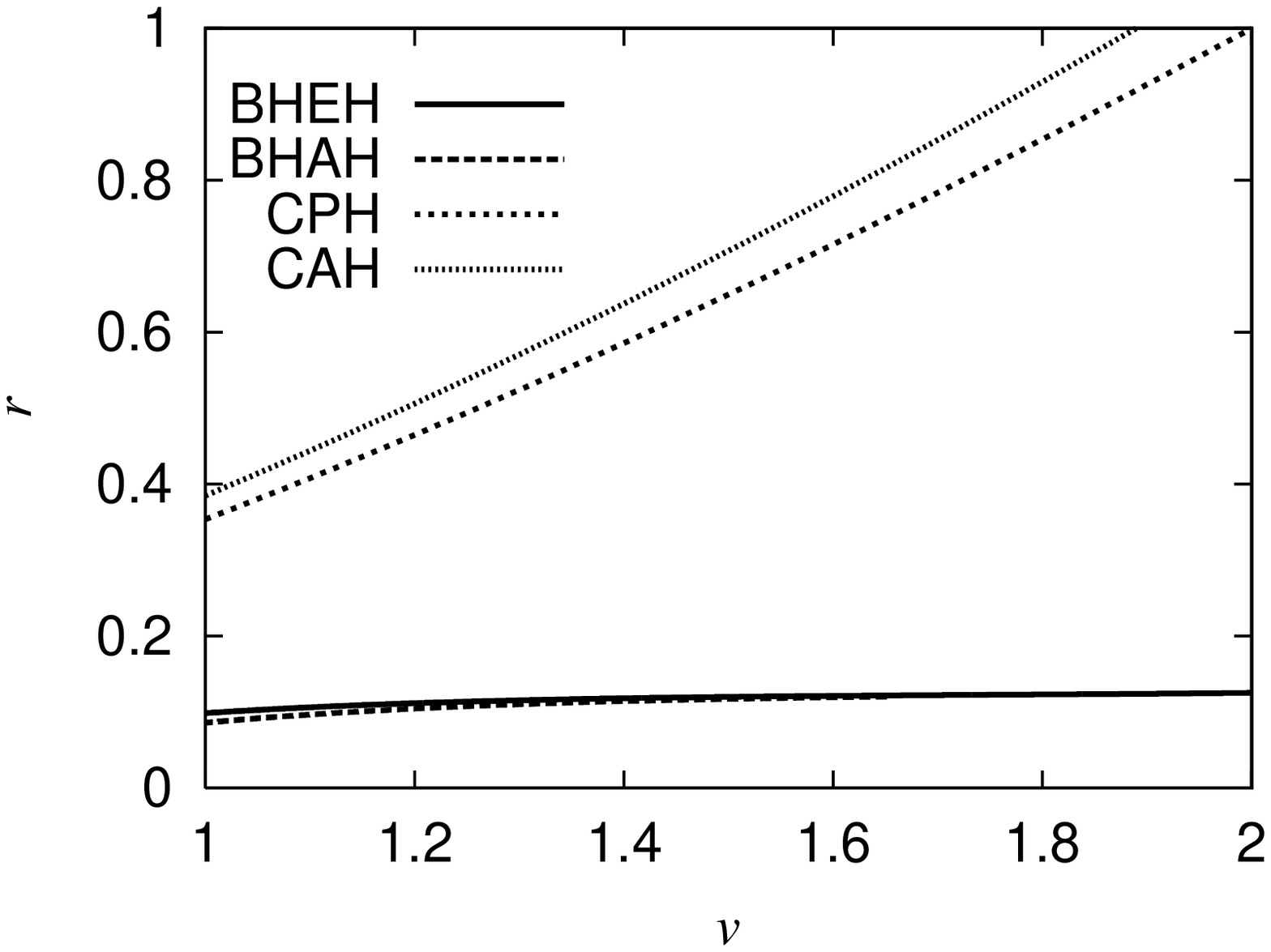}}\\
\subfigure[]{\includegraphics[scale=0.45]{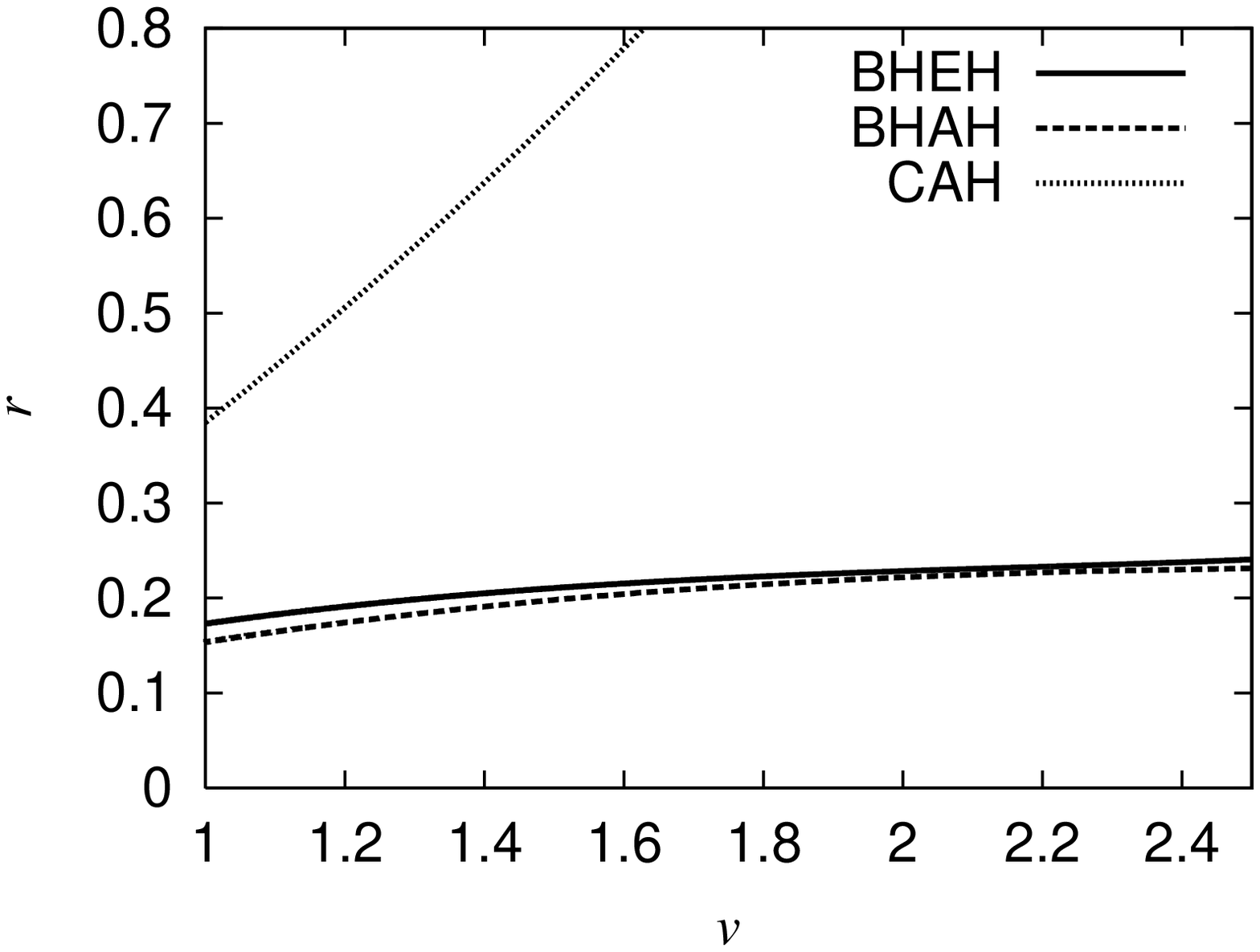}}
&\subfigure[]{\includegraphics[scale=0.45]{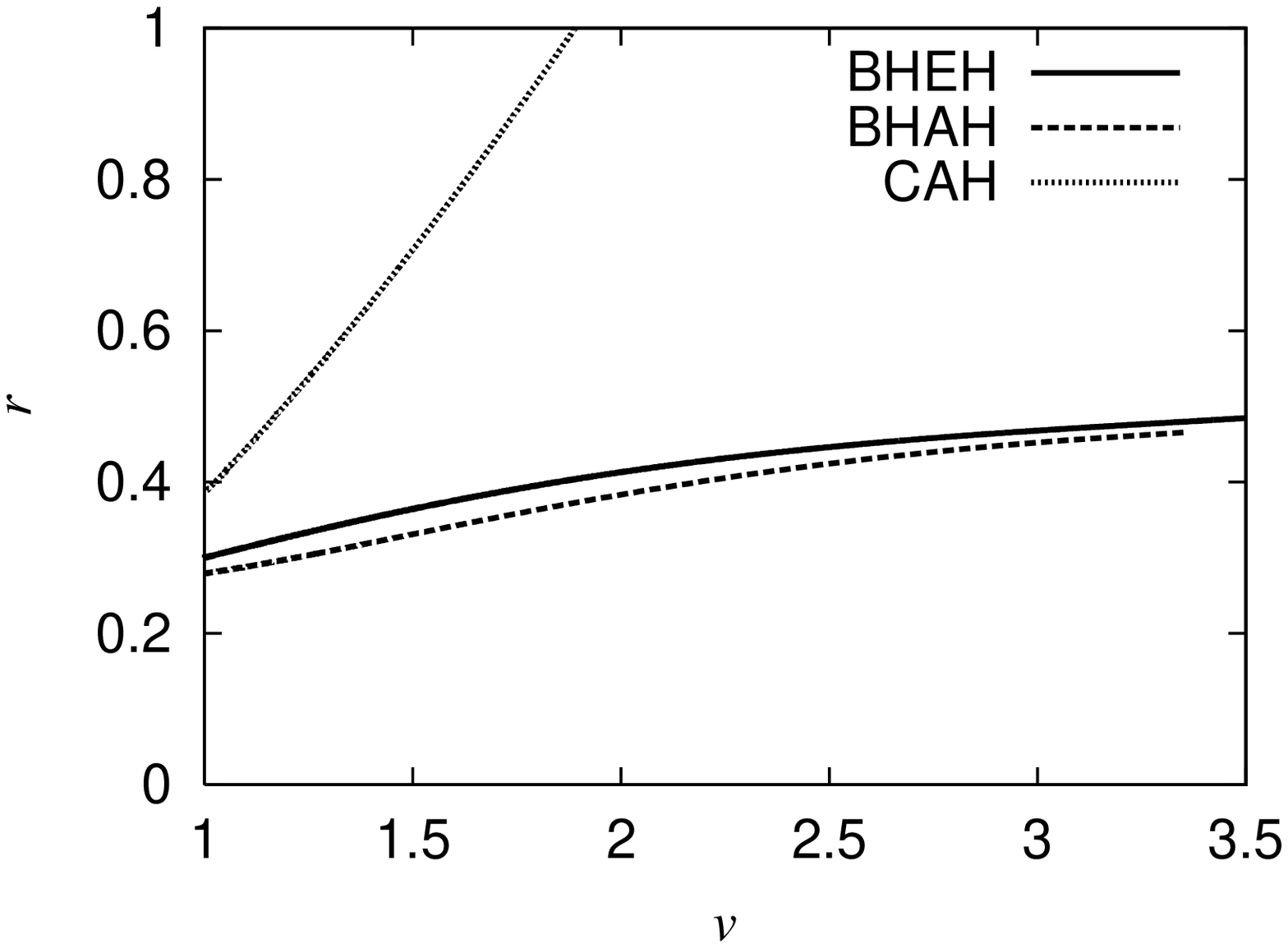}}\\
\end{tabular}
\caption{\label{fig:eh_rad}
The evolution of the area radius $r$ of the black hole 
event horizon, black hole apparent horizon and 
cosmological apparent horizon for Models A--D
is plotted in (a)--(d).
For Model B, the radius of the cosmological particle horizon
of the unperturbed flat Friedmann solution is also plotted.}
\end{figure}

\begin{figure}[htbp]
\begin{tabular}{cc}
\subfigure[]{\includegraphics[scale=0.45]{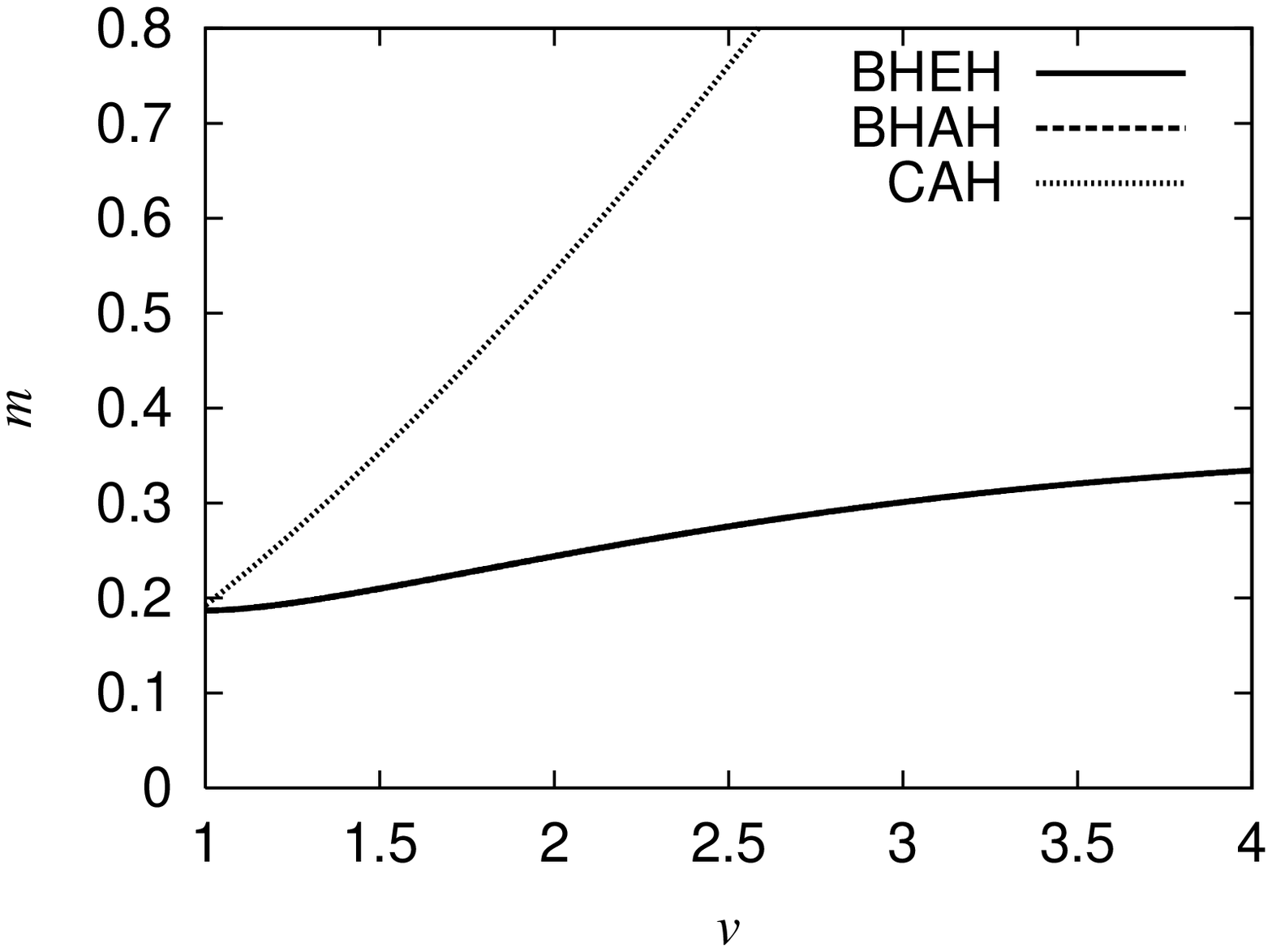}}&
\subfigure[]{\includegraphics[scale=0.45]{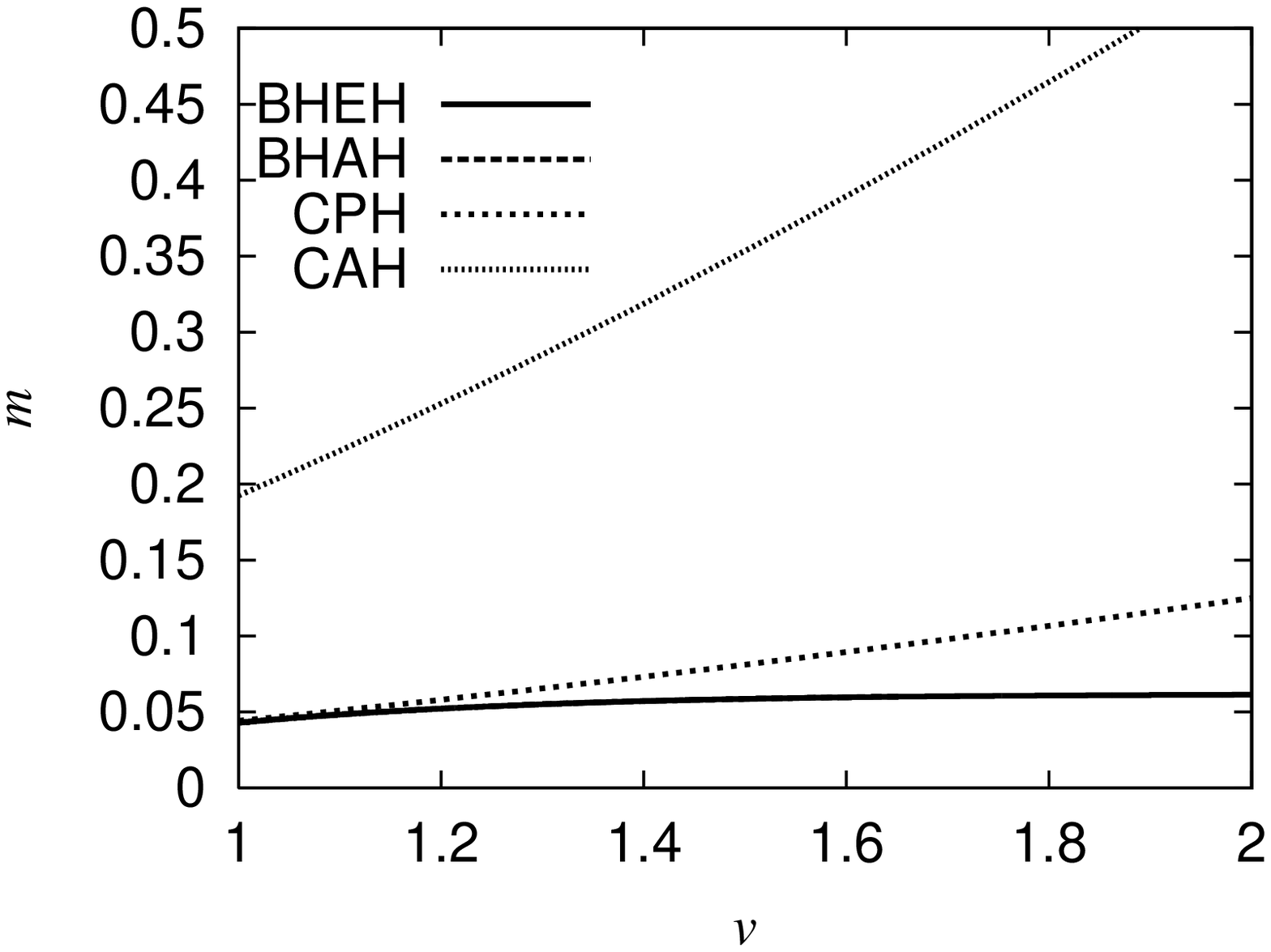}}\\
\subfigure[]{\includegraphics[scale=0.45]{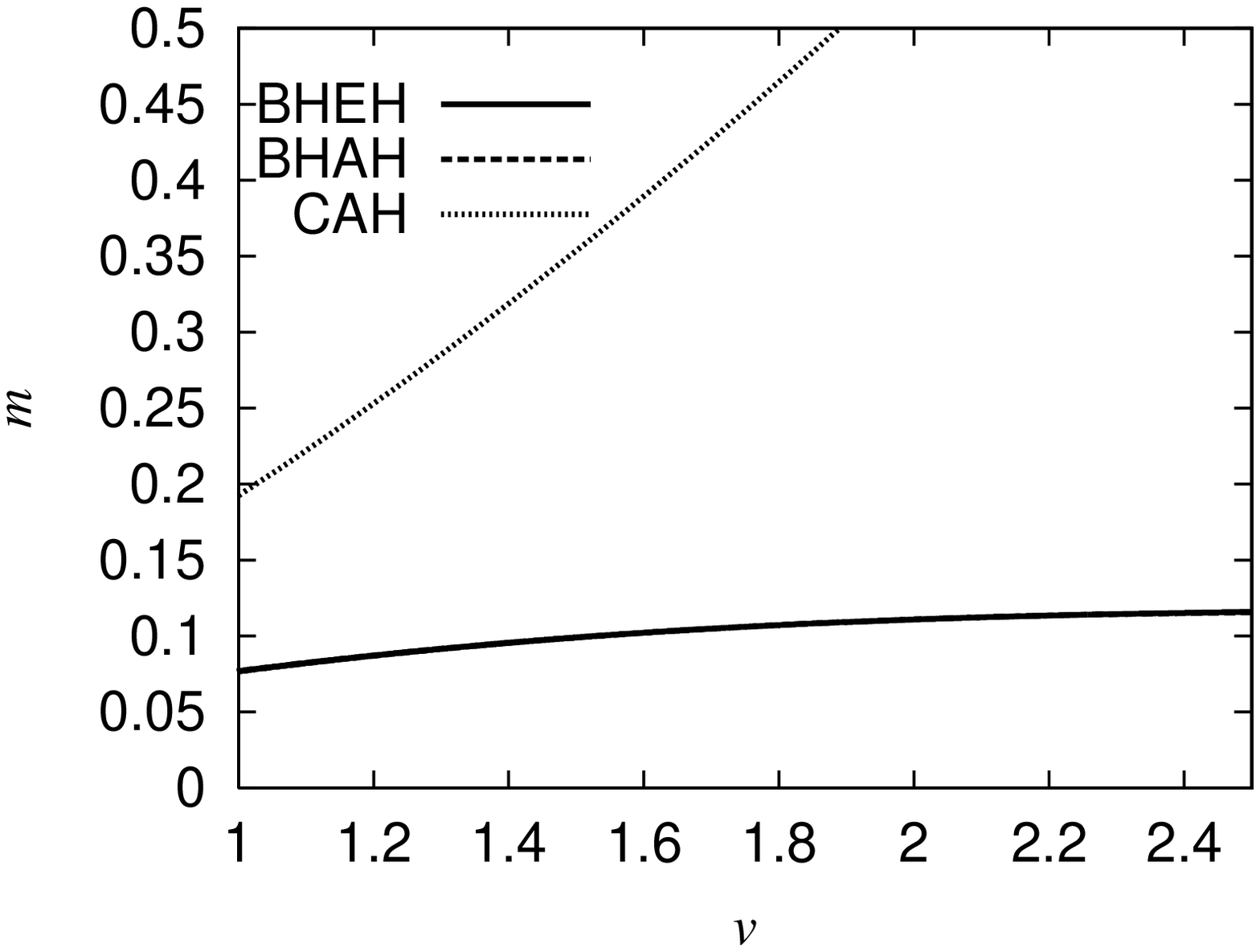}}
&\subfigure[]{\includegraphics[scale=0.45]{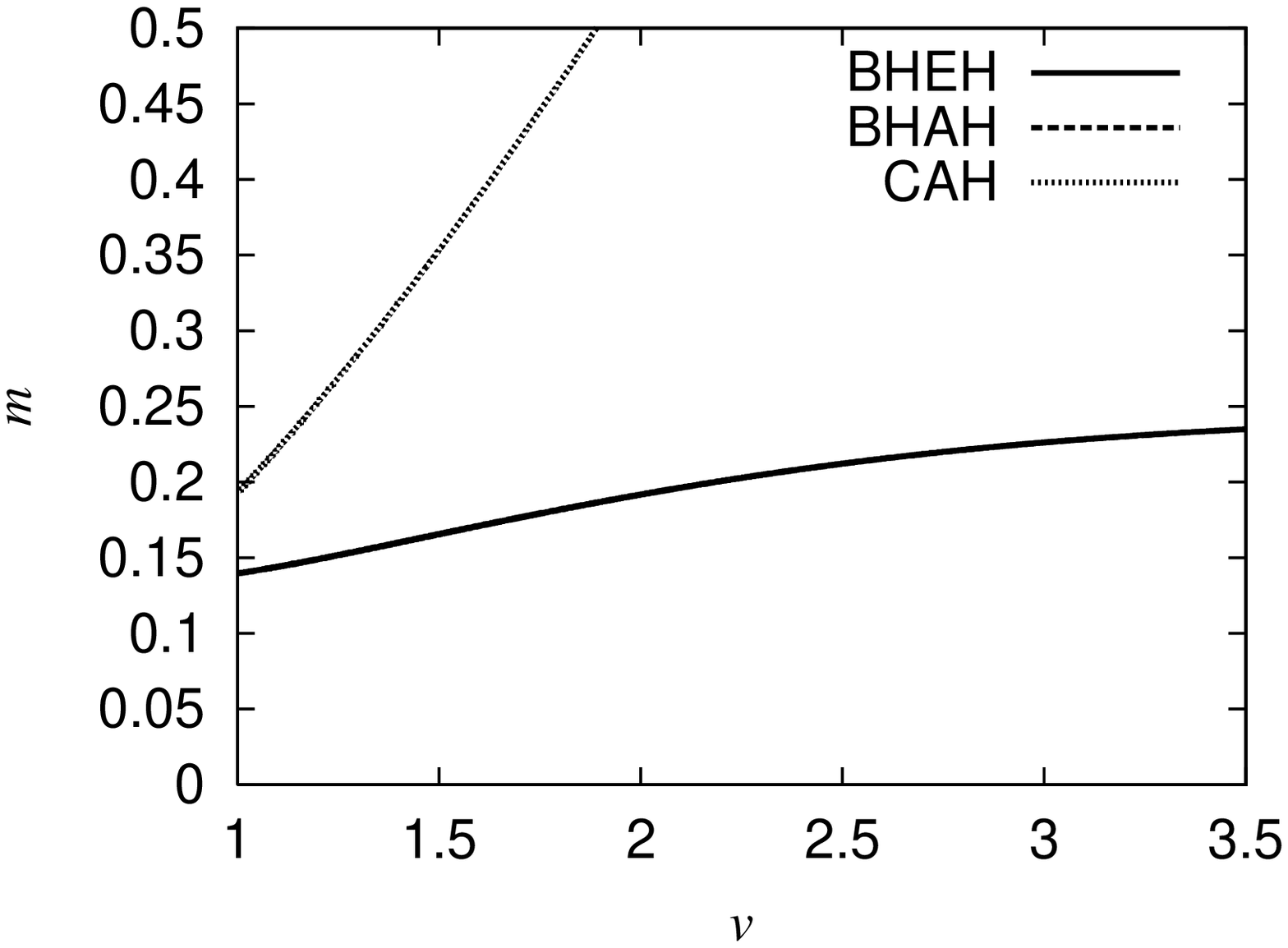}}\\
\end{tabular}
\caption{\label{fig:eh_mass}
The evolution of the mass $m$ contained within the black hole 
event horizon, black hole apparent horizon 
and cosmological apparent horizon for Models A--D
is plotted in (a)--(d).
The curves for the black hole event horizon and 
the black hole apparent horizon are 
indistinguishable.
For Model B, the mass contained within the cosmological particle horizon
of the unperturbed flat Friedmann solution is also plotted.}
\end{figure}

The radii and masses of the black hole apparent horizon
and event horizon for Models A--D are shown, together with those for the 
cosmological horizons,
in Figs.~\ref{fig:eh_rad} and \ref{fig:eh_mass}.
The initial and final values and their ratios
are also summarised in Tables~\ref{tb:radii} and \ref{tb:masses}.
All three horizons approximately coincide
at $v=1$ for Model A, both in radius and mass,
as seen in Figs.~\ref{fig:eh_rad}(a) and ~\ref{fig:eh_mass}(a). 
Figs.~\ref{fig:eh_rad} and ~\ref{fig:eh_mass} 
show that the qualitative features of the evolution for 
Models A--D are very similar. The radius of the cosmological
apparent horizon of the unperturbed 
flat Friedmann solution grows like $v^{3/2}$,
while the black hole apparent horizon and event horizon
converge and grow much more slowly.
As a consequence, the black hole
horizons soon get much smaller than the cosmological apparent 
horizon. This is also the case for the corresponding
masses. 

\begin{figure}[htbp]
\begin{tabular}{cc}
\subfigure[]{\includegraphics[scale=0.45]{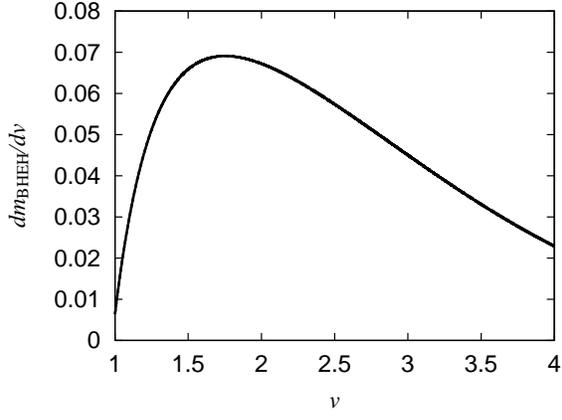}} &
\subfigure[]{\includegraphics[scale=0.45]{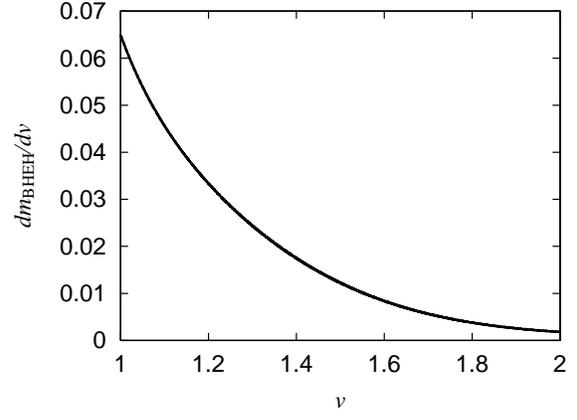}} \\
\subfigure[]{\includegraphics[scale=0.45]{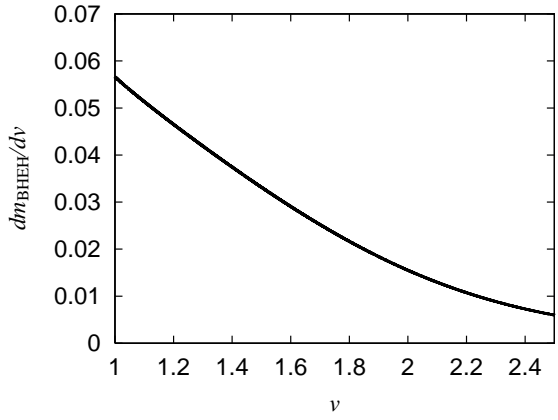}} 
&\subfigure[]{\includegraphics[scale=0.45]{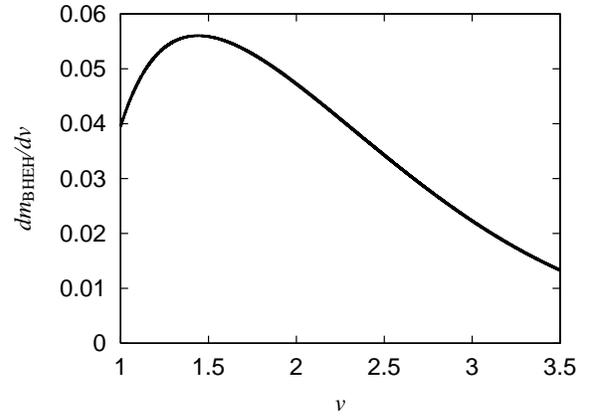}} \\
\end{tabular}
\caption{\label{fig:mass_accretion}
The PBH mass accretion rate $dm_{\rm BHEH}/dv$ 
for Models
A--D is plotted in (a)--(d).}
\end{figure}

Since the black hole event horizon is given by 
the curve $u=u_{\rm BHEH}=\mbox{const}$, Eq.~(\ref{eq:m_v})
gives the mass accretion rate of the black hole.
This also serves as a consistency check 
for numerical accuracy.
The result for Models A--D is shown in Fig.~\ref{fig:mass_accretion}, where
both 
sides of Eq.~(\ref{eq:m_v}) are plotted.
We find that the curves are almost indistinguishable,
which implies very good numerical accuracy.
For Model A, the accretion rate starts very small,  
then increases, reaches a maximum of
$ 0.0691$ at $v\simeq 1.75$ and
then decreases to a very small value.
Recall that the units are such that
the background Hubble parameter is 4 and 
the background Hubble length is 0.25 at $u_{0}$ and $v_{0}$.
For Models B and C, the accretion rate 
starts with its maximum value and then 
monotonically decreases.
The features for Model D are similar to those for Model A. 

An important qualitative feature is that 
accretion is suppressed for a PBH 
nearly as large as the cosmological apparent horizon. 
(In Model D the black hole event horizon happens 
to be close to the cosmological 
apparent horizon.) This is because Eq.~(\ref{eq:m_v}) implies that
the mass accretion rate is proportional to $(-r_{,u})$,
 i.e. the ingoing null contraction,
and this vanishes at the cosmological 
apparent horizon.
This feature is a purely general relativistic effect and there is no
analogue in Newtonian gravity. It can be understood physically as
arising because the scalar field must be partaking in the cosmic expansion at
sufficiently large distances, rather than accreting onto the black
hole. Since the sound speed is the speed of light for a scalar field, the cosmological apparent horizon is likely to be the transition point. A more rigorous explanation of this effect is presented in Section V, where it is shown to apply for more general fluids. 

We conclude that, even if a PBH starts off
as large as the cosmological apparent horizon, 
it soon gets considerably smaller than it.
Whatever the initial size of the PBH, accretion can be large only
for about a Hubble time, after which it
becomes ineffective.
As seen in Tables~\ref{tb:radii} and \ref{tb:masses}, 
the mass increase is at most 79 \% for our models. 

\subsection{Geometry and scalar field}
\begin{figure}[htbp]
\begin{tabular}{cc}
\subfigure[]{\includegraphics[scale=0.45]{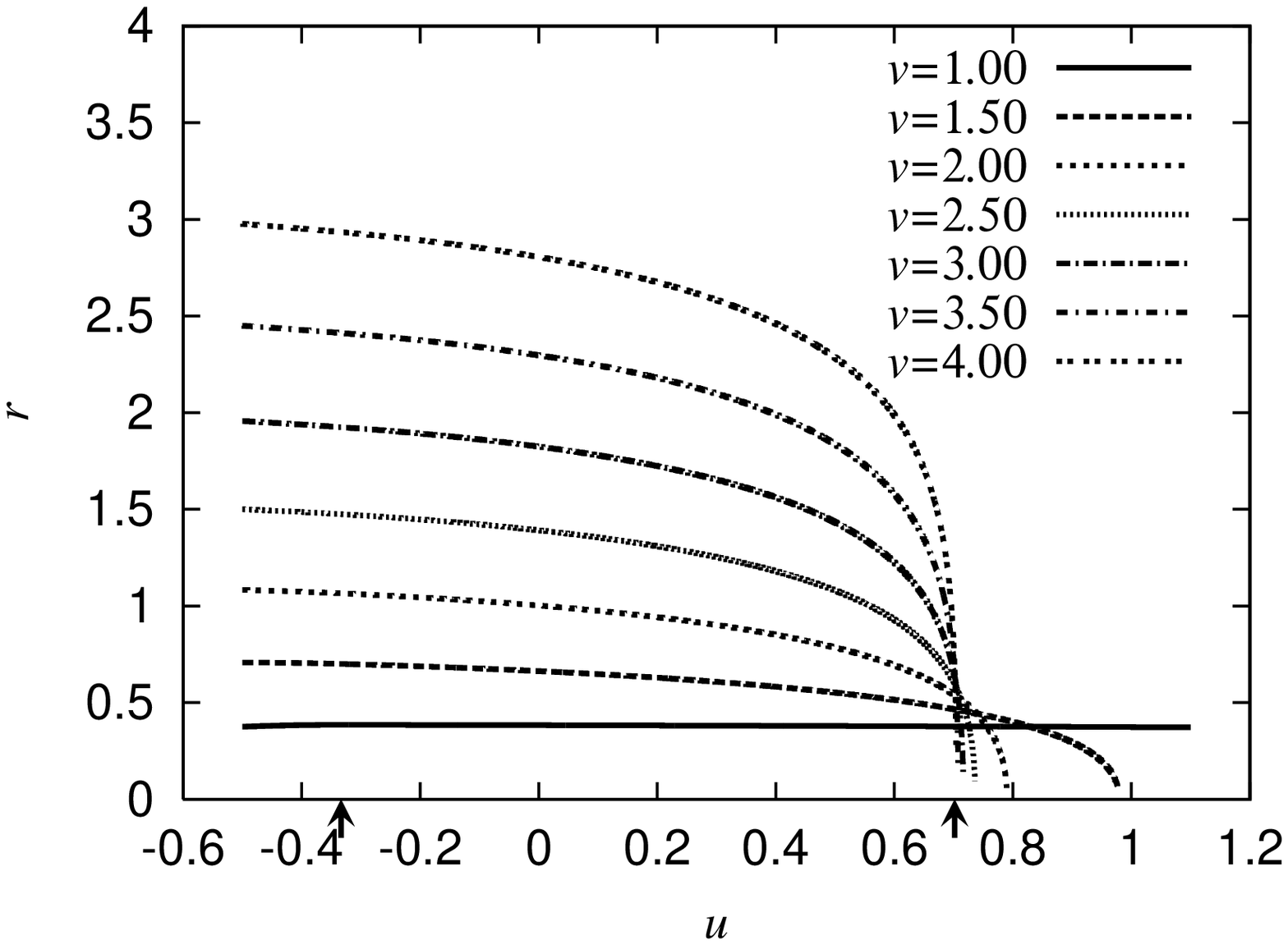}}&
\subfigure[]{\includegraphics[scale=0.45]{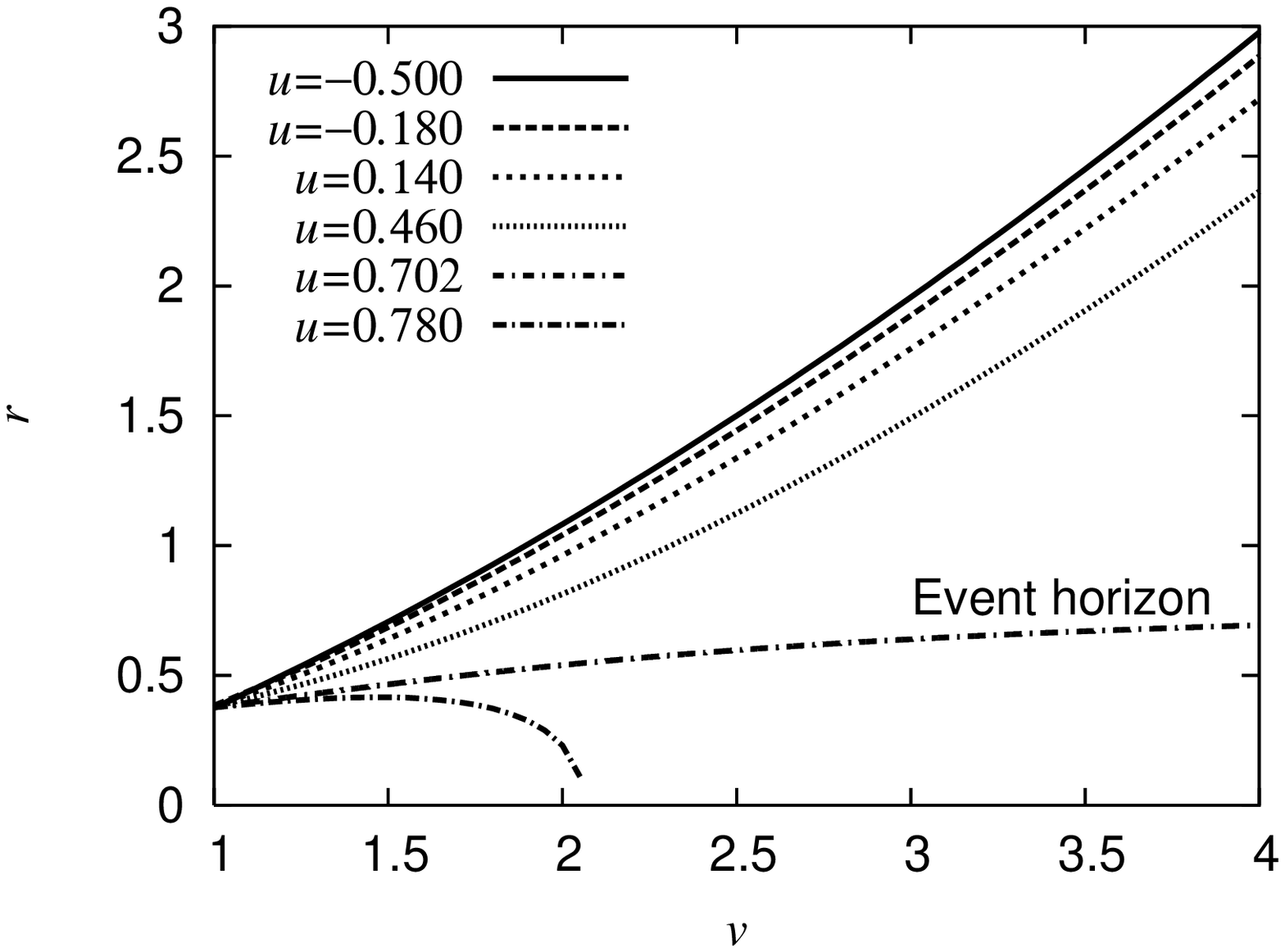}}
\end{tabular}
\caption{\label{fig:rad}
The area radius $r$ along (a) ingoing null rays $v=\mbox{const}$
and (b) outgoing null rays $u=\mbox{const}$ for Model A.
The arrows at $u=u_{\rm m}=-1/3$ and $u=u_{\rm BHEH}\simeq 0.702$ 
in (a) denote the matching outgoing null surface and 
black hole event horizon, respectively.}
\end{figure}

Since the qualitative features of the numerical results 
are very similar for Models A--D,
in this subsection we concentrate on Model A.
Figure~\ref{fig:rad}
shows the evolution of the area radius $r$ in terms of $u$ and $v$. 
We can see from Fig.~\ref{fig:rad}(b)
that there is a threshold value of $u$:
outgoing null rays with smaller $u$
go to infinity, while those with larger $u$
go to the singularity at $r=0$.
Therefore, this threshold value can be identified 
with the black hole event horizon $u_{\rm BHEH}\simeq 0.702$.
Since $r_{,v}$ approaches zero
along the black hole event horizon, 
all ingoing null rays seem to cross it
at almost the same radius in Fig.~\ref{fig:rad}(a).
In fact, the integration is continued well beyond $v=v_{1}$
to locate the black hole event horizon accurately.
Figure~\ref{fig:rad}(b) shows that the radius of the black hole event
horizon reaches $0.694$ at $v=4$ and this might be regarded as its asymptotic value.

\begin{figure}[htbp]
\begin{tabular}{cc}
\subfigure[]{\includegraphics[scale=0.45]{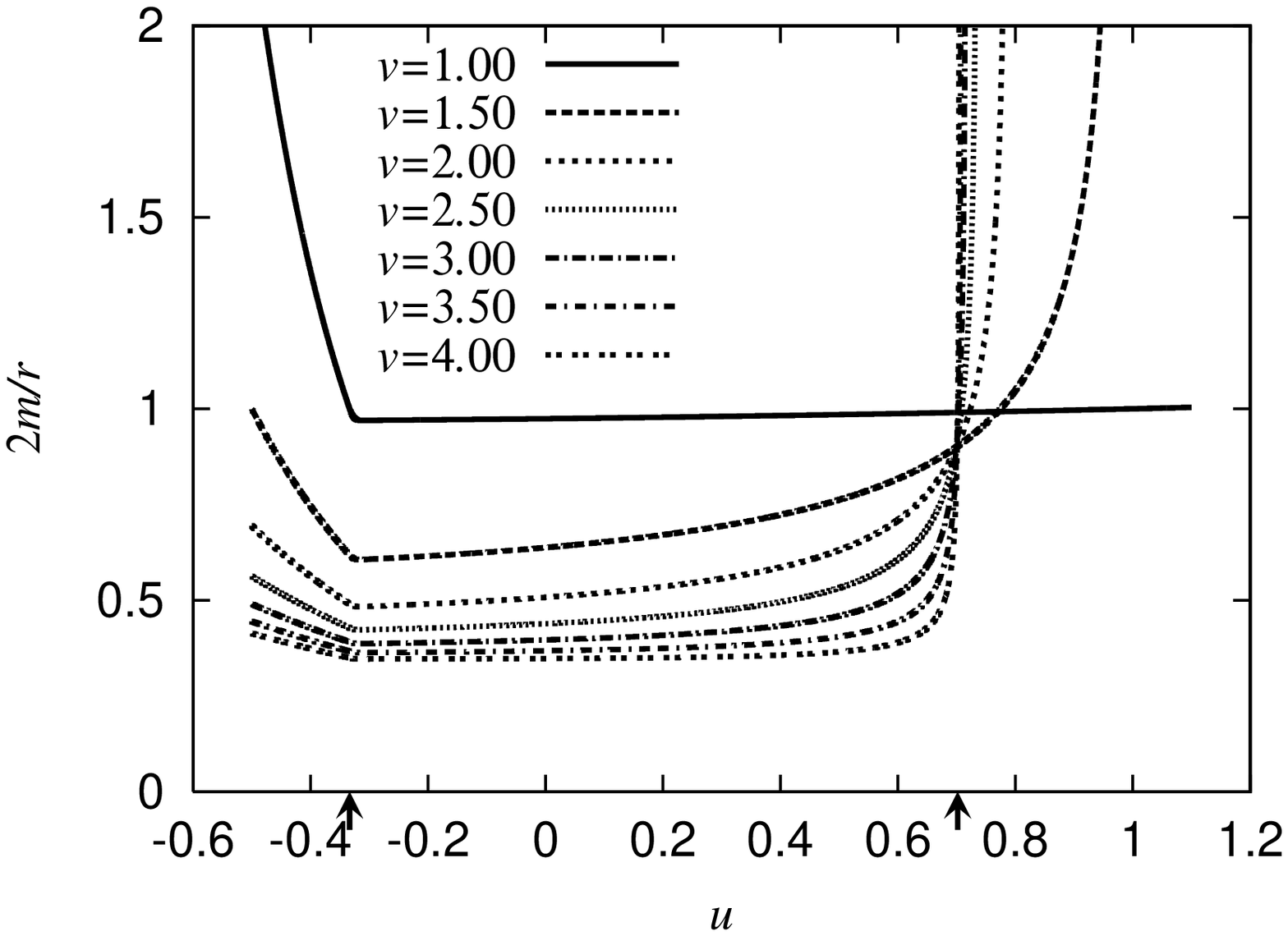}}&
\subfigure[]{\includegraphics[scale=0.45]{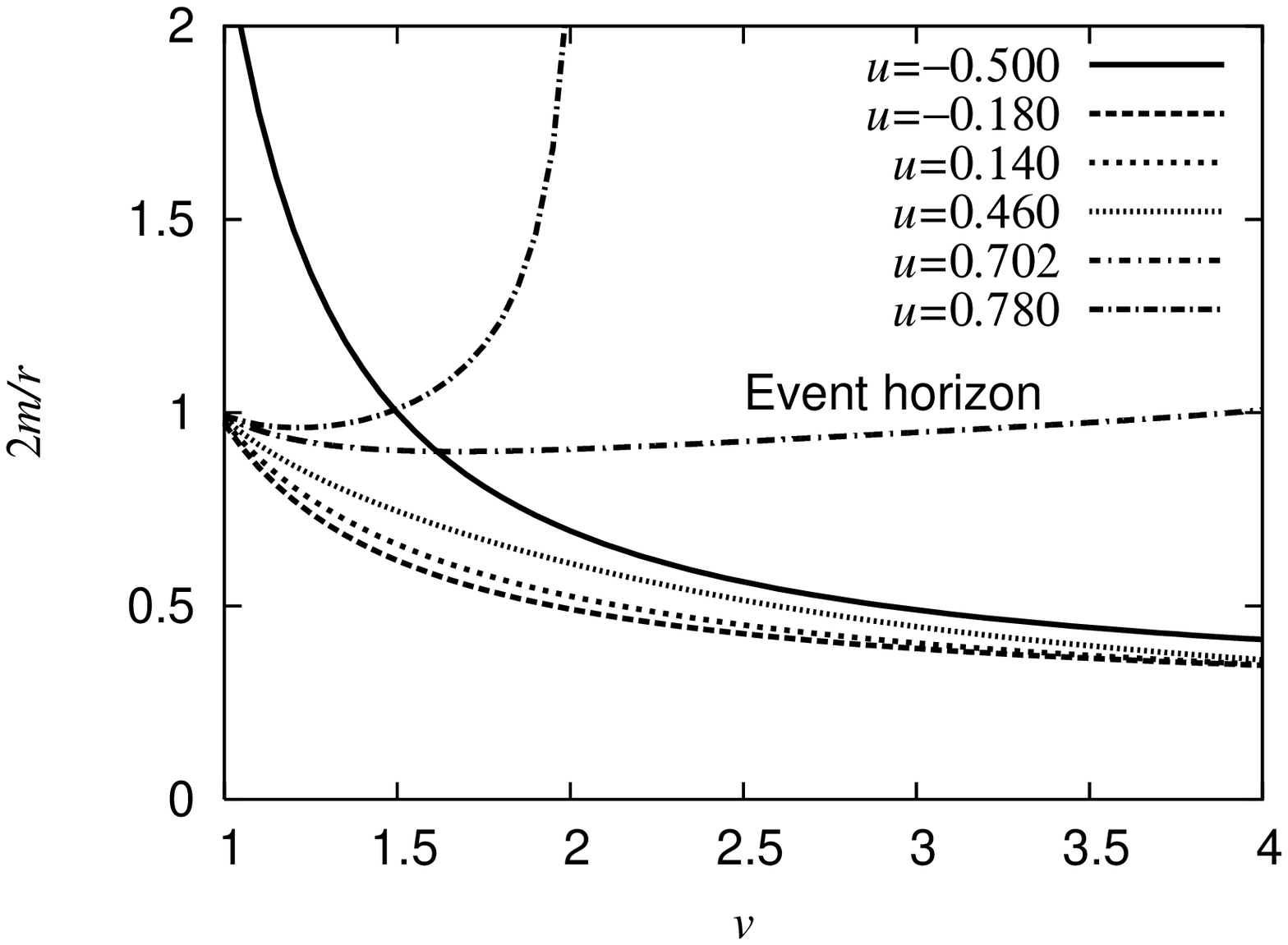}}
\end{tabular}
\caption{\label{fig:pot}
Same as Fig.~\ref{fig:rad} but for $2m/r$.}
\end{figure}

Figure~\ref{fig:pot} shows the evolution of $2m/r$ in terms of $u$ and $v$.
There is an apparent horizon where this is unity.
The signs of both ingoing and outgoing null expansions
are the same when this is larger than unity.
A region is untrapped when it is smaller than unity.
As seen in Fig.~\ref{fig:pot}(a), $2m/r$ is almost exactly unity 
in the perturbed region at the initial ingoing null surface $v=v_{0}=1$.
As $u$ increases, there then appears a black hole apparent horizon, 
beyond which the region is trapped. 
There is also a cosmological apparent
horizon at $u=-1/3$ and $v=v_{0}=1$, 
although it goes outside the calculated region 
during the evolution. As time proceeds, the value of $2m/r$ in the 
untrapped region between the black hole and cosmological 
apparent horizons 
becomes smaller and the curve near the black hole apparent horizon
becomes steeper. As seen in Fig.~\ref{fig:pot}(b),
there is no black hole apparent horizon before the event horizon
forms at $u=u_{\rm BHEH} \simeq 0.702$.
At $u=u_{\rm BHEH}$, $2m/r$ approaches unity from below
as $v$ increases.
Soon after $u=u_{\rm BHEH}$, a black hole 
apparent horizon appears.

\begin{figure}[htbp]
\begin{tabular}{cc}
\subfigure[]{\includegraphics[scale=0.45]{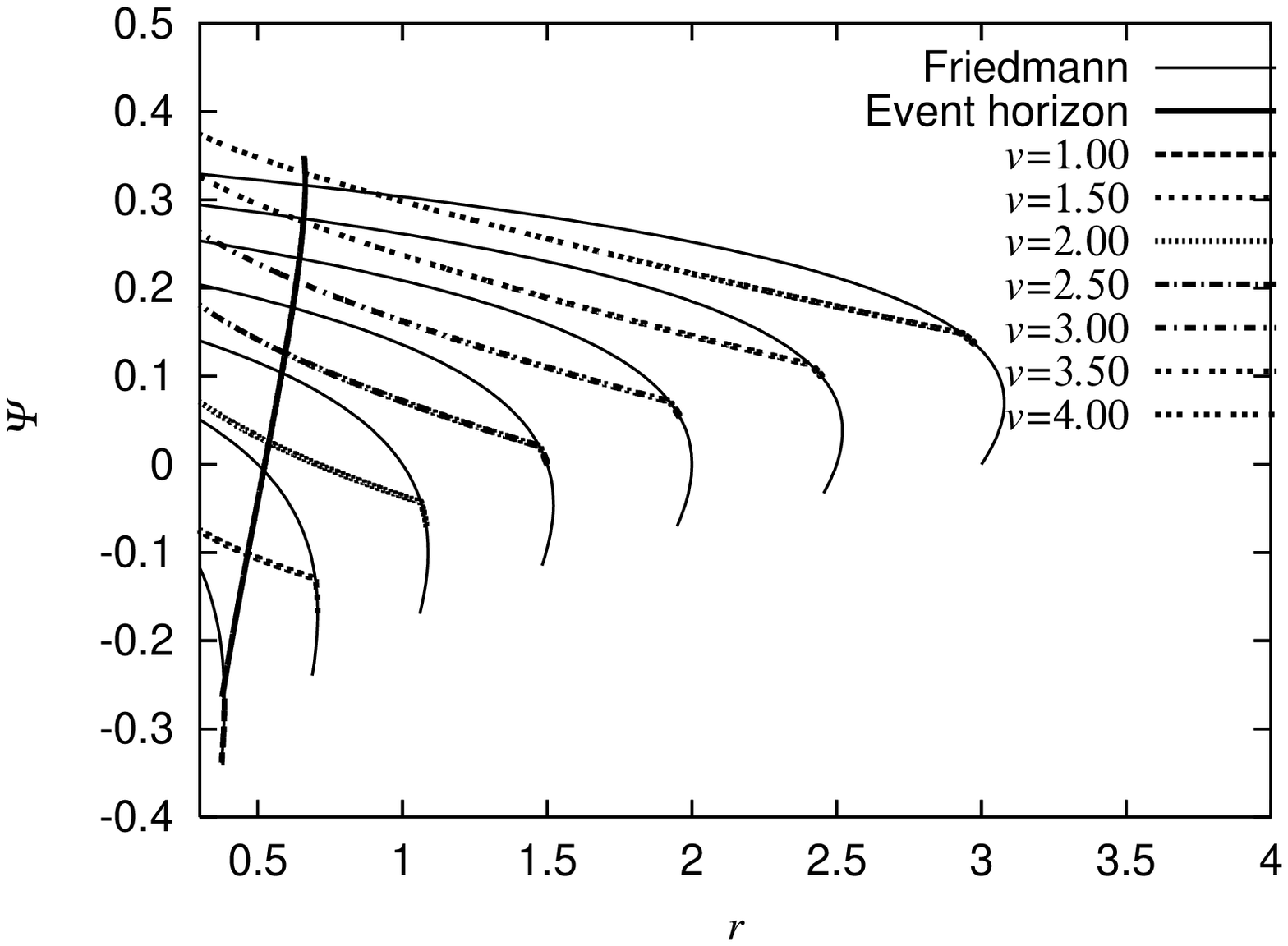}}&
\subfigure[]{\includegraphics[scale=0.45]{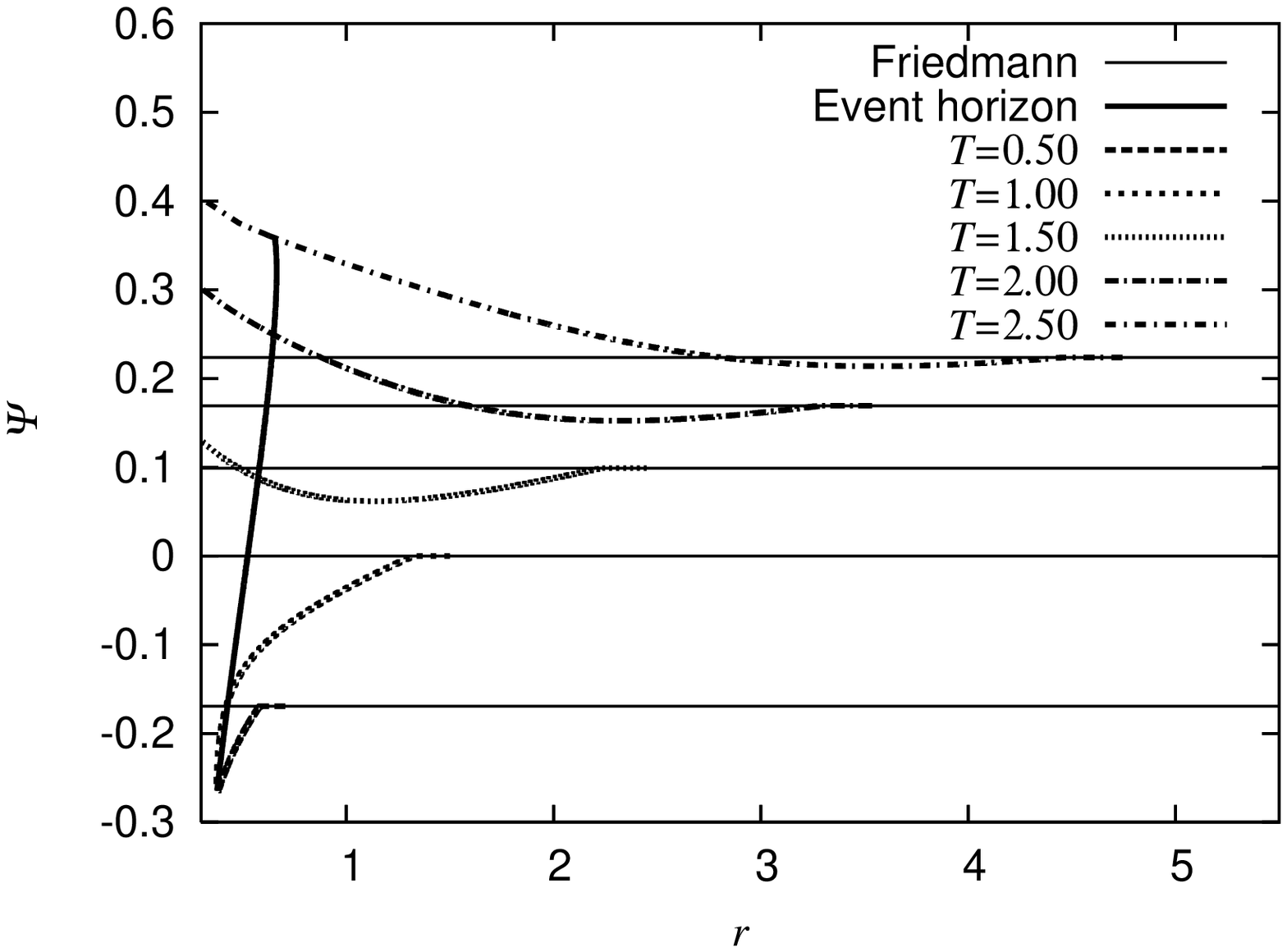}}
\end{tabular}
\caption{\label{fig:scalar_compare}
The snapshots of scalar field profiles for Model A
on a sequence of hypersurfaces with (a) constant $v$
and (b) constant $T$.
The profiles are indicated by broken curves for the 
numerical simulations and by solid light curves
for the flat Friedmann solution. The heavy solid 
curves give the position of the event horizon.
}
\end{figure}

Figure~\ref{fig:scalar_compare} shows the evolution 
of the scalar field for Model A.
For clarity, it is plotted in terms of both $(v,r)$ and $(T,r)$
coordinates. In either case there is no coordinate singularity 
at the black hole horizons as the ingoing Eddington-Finkelstein
coordinate is well-behaved 
for the Schwarzschild black hole.
In the $(v,r)$ diagram of Fig.~\ref{fig:scalar_compare}(a), 
the broken and solid light curves give 
the results of the simulations and 
the evolution of the flat Friedmann solution,
respectively.
Because of the initial perturbation, the value of the 
scalar field in the perturbed region 
starts off much smaller than that of the flat Friedmann 
solution.
Thereafter, the perturbed scalar field 
tends to evolve as in
the flat Friedmann solution.
The position of the event horizon is shown by the solid heavy curve.
Since the radius of the black hole event horizon
is $0.694$ at $v=4$,
the region inside the black hole event horizon 
is also calculated accurately.
The evolution of 
the scalar field is well described by the 
flat Friedmann evolution with a small perturbation 
both outside and inside the black hole event horizon. 
The scalar field is smooth at the black hole event horizon.
Fig.~\ref{fig:scalar_compare}(b) shows
the equivalent results in the $(T,r)$ diagram.
It should be stressed,
however, that there is no unique choice of spatial 
hypersurface and the profile of $\Psi$
would be different with another choice.

\begin{figure}[htbp]
\begin{tabular}{cc}
\subfigure[]{\includegraphics[scale=0.45]{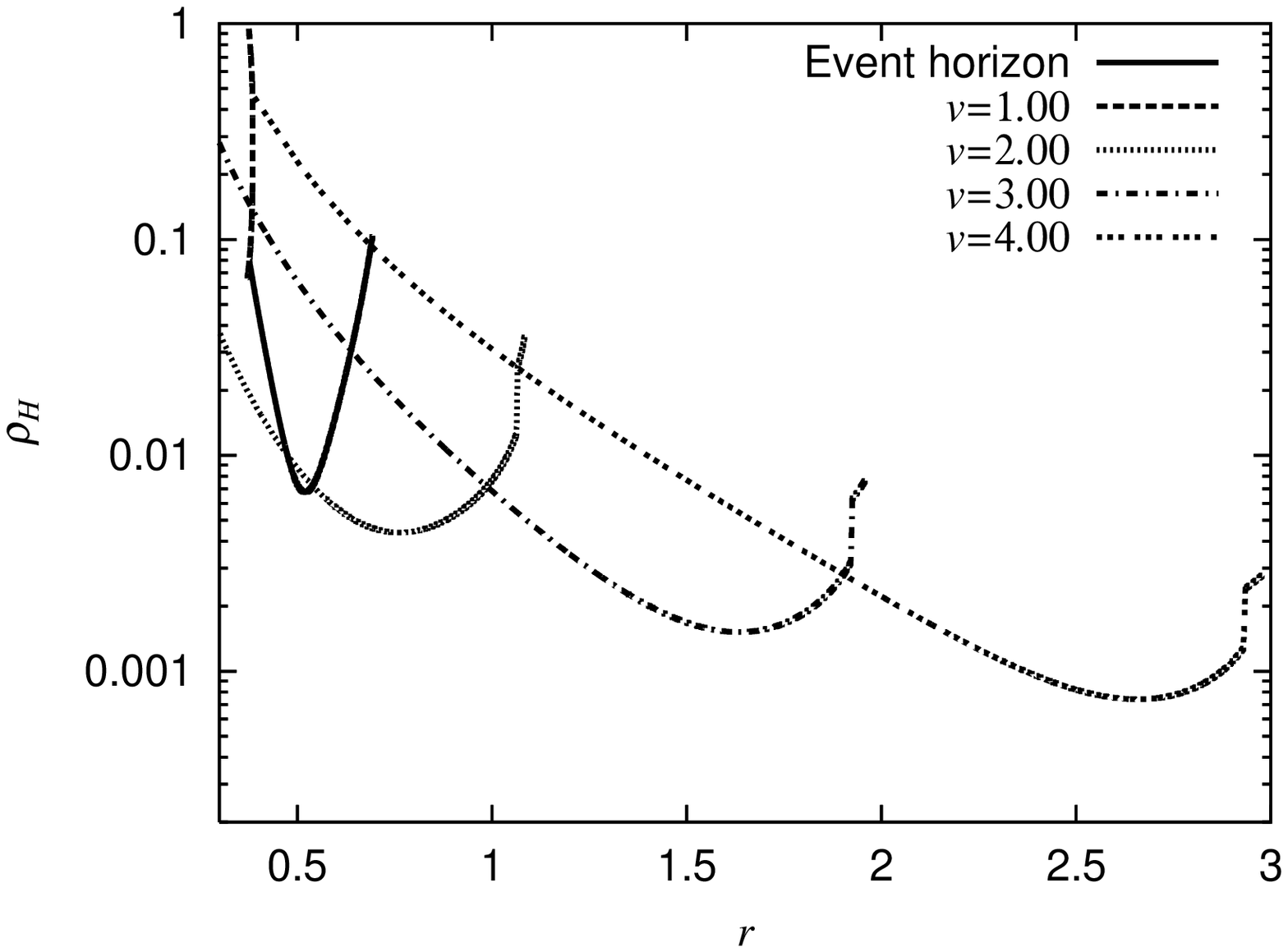}}&
\subfigure[]{\includegraphics[scale=0.45]{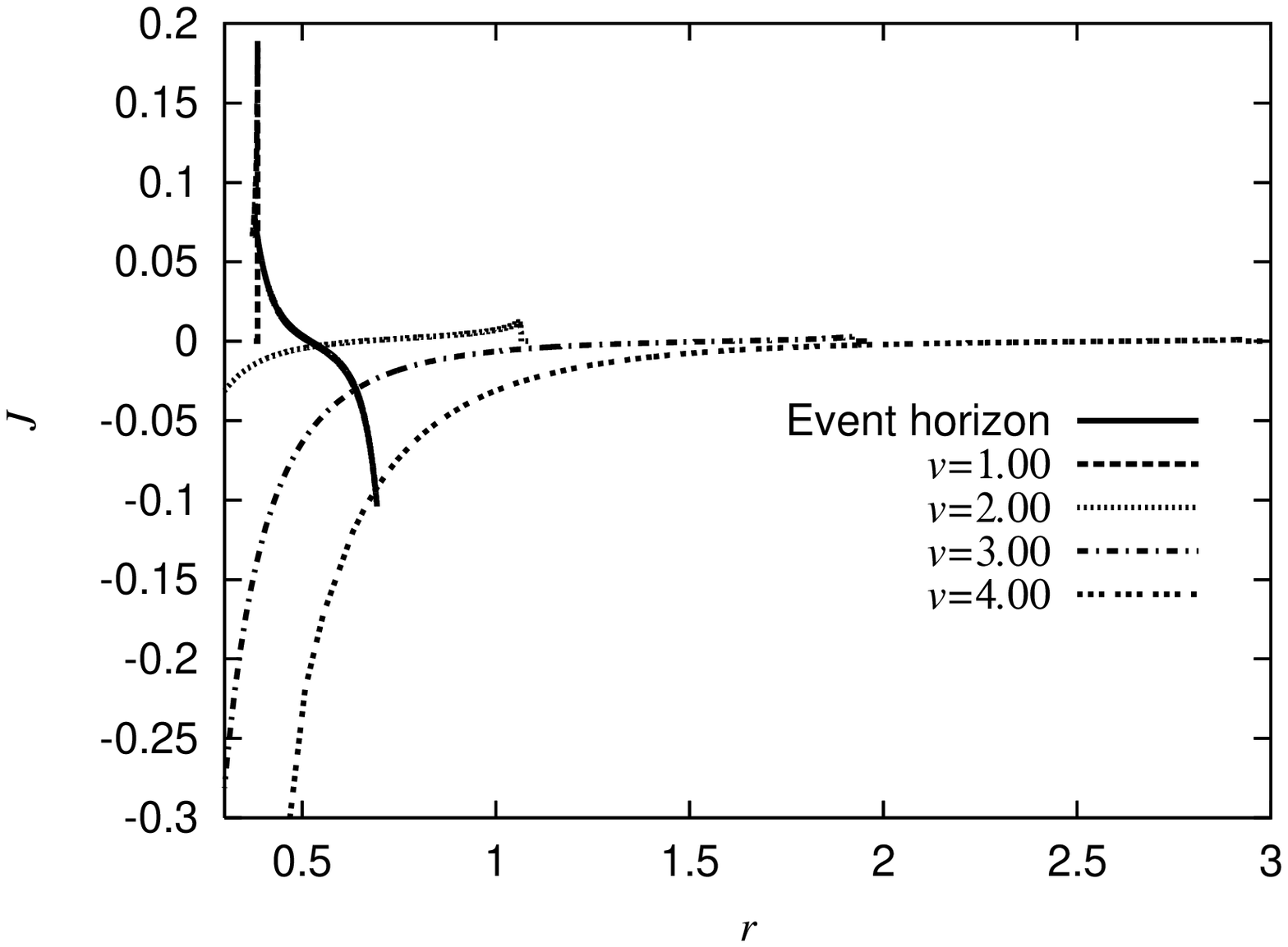}}
\end{tabular}
\caption{\label{fig:3+1}
The snapshots of the energy density $\rho_{\rm H}$ 
and the momentum density $J$ of the scalar field measured 
by the observer moving normal to the $T=(u+v)/2=\mbox{const}$ spacelike
hypersurface for Model A.
Note that the event horizon expands with respect to $r$ as time proceeds.}
\end{figure}

Figure~\ref{fig:3+1} shows the energy density $\rho_{\rm H}$
and momentum density $J$ measured by the observer moving normal to the 
$T=\mbox{const}$ spacelike hypersurface for Model A.
Since this normal observer coincides with the comoving 
observer in the flat Friedmann universe,
$J$ vanishes in the flat Friedmann region for $u<u_{\rm m}$
(see Appendix~\ref{subsec:flat_friedmann}).
In terms of the energy density $\rho_{\rm H}$, 
there is an underdense region inside the 
flat Friedmann region. Inside this underdense region, 
there is an overdense region around 
the black hole event horizon, where the energy density increases with time. 
In terms of the momentum density $J$,
there is considerable energy infall around
the black hole event horizon.
$|J|$ increases significantly near the black hole event horizon,
although we should note that this is an observer-dependent view.

\begin{figure}[htbp]
\begin{tabular}{cc}
\subfigure[]{\includegraphics[scale=0.45]{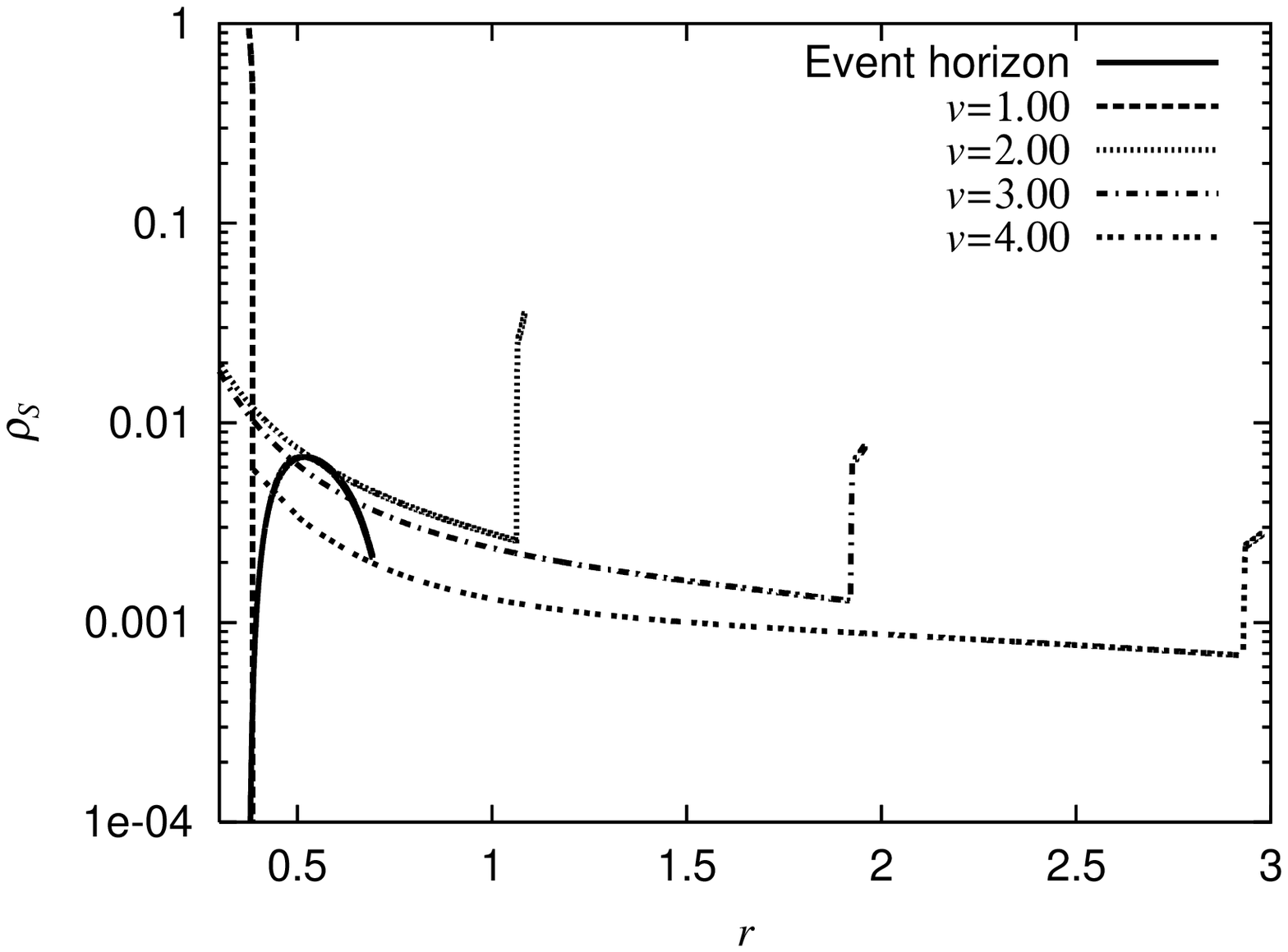}}&
\subfigure[]{\includegraphics[scale=0.45]{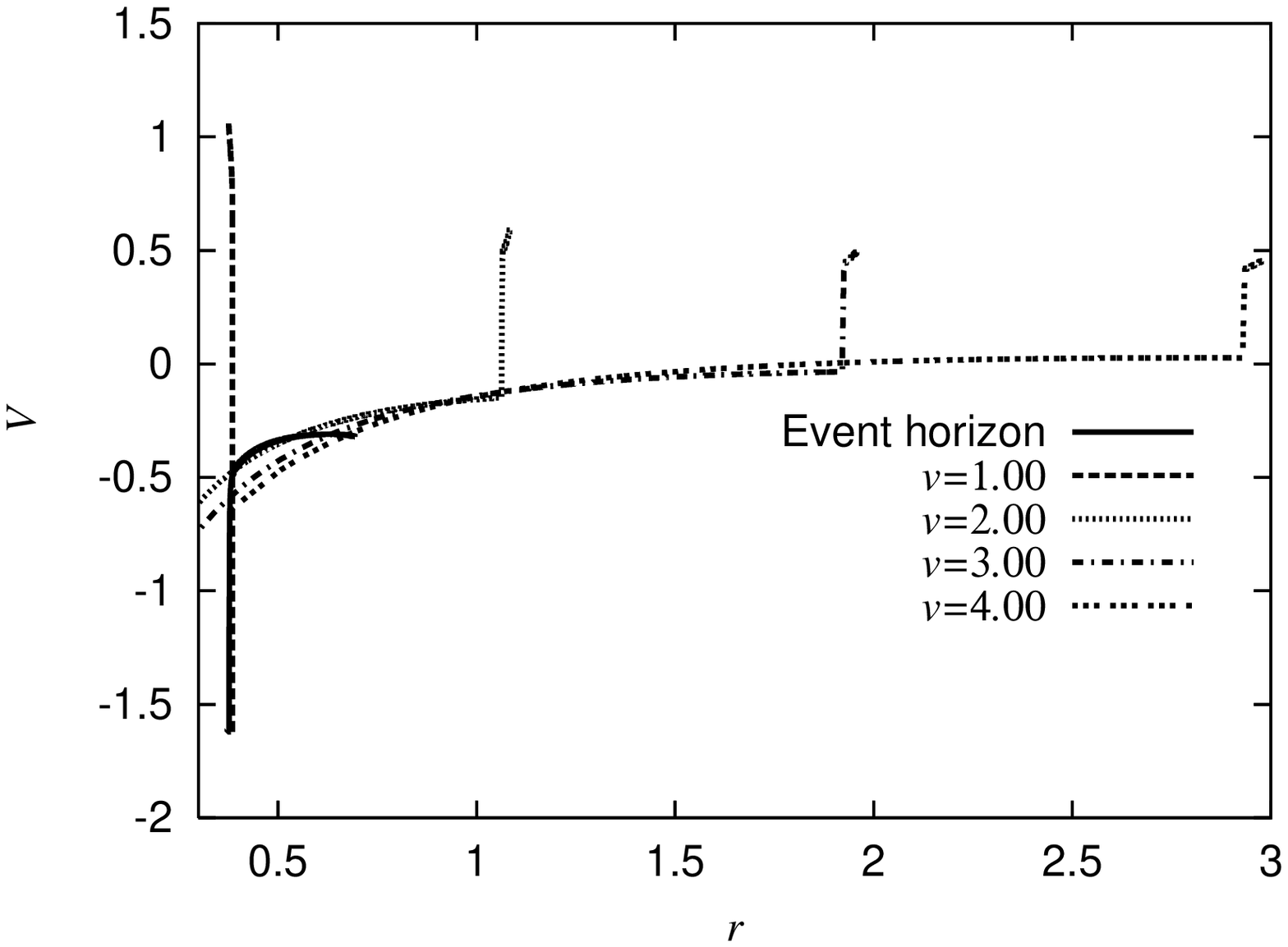}}
\end{tabular}
\caption{\label{fig:sf}
The snapshots of (a) the energy density $\rho_{\rm S}$ and (b)
the velocity $V$ of the stiff fluid equivalent to the 
scalar field for Model A. It is noted that 
both $\rho_{\rm S}$ and $V$ are 
observer-independent.
The event horizon always expands as time proceeds.}
\end{figure}

Figure~\ref{fig:sf} shows the
observer-independent quantities $\rho_{\rm S}$ and $V$. These are the energy density
and velocity of the equivalent stiff fluid. The spikes correspond to the matching coordinate $u_m$.
It is interesting that $\rho_{\rm S}$ is everywhere positive.
This is a non-trivial result since $\rho_{\rm S}$ could in principle be negative if the spatial gradients were large enough. This is related to the fact, as illustrated in Fig.~\ref{fig:sf}(a), that the gradient of the scalar field is
always timelike in this simulation. (Otherwise it would not be equivalent to a stiff fluid.)
Also, in terms of the density $\rho_{\rm S}$, we can see an underdense region 
just inside the flat Friedmann region. Within this underdense region,
the energy density increases as one moves
inwards. However, the energy density decreases 
nearly homogeneously for $v\ge 2.5$.
In terms of the velocity $V$,
the region near the black hole is infalling, as 
seen in Fig.~\ref{fig:sf}(b),
but the infall velocity profile near the black hole event horizon
is almost stationary.
We conclude that the growth 
of $\rho_{\rm H}$ and $|J|$
around the black hole event horizon is completely due to 
the Lorentz factor.
Physically, the energy density $\rho_{\rm S}$ is slowly decreasing 
and the infall velocity profile $V$ is almost stationary
near the black hole event horizon.
These results suggest that
the PBH can be regarded as almost 
isolated, with a relatively small mass accretion
for $v\agt 2$.
This is consistent with the mass accretion rate seen 
in Fig.~\ref{fig:mass_accretion}(a).
Although there are apparent discontinuities
at large radii in Figs.~\ref{fig:3+1} and 
\ref{fig:sf}, this is only due to the narrow boundary
and all quantities are continuous everywhere.

\section{Discussion}
\subsection{Self-similar growth and the Newtonian accretion formula}

Various arguments~\cite{zn1967,ch1974,cg1999,hc2004a} lead to an estimate 
for the mass accretion rate of a PBH in a cosmological background 
of the form
\begin{equation}
\frac{dm_{\rm BH}}{dt}=4 \pi \alpha \rho v_{\rm s} r_{\rm A}^2 
\label{eq:newt_accretion}
\end{equation}
where $\rho$ is the background density, $v_{\rm s}$ is the sound speed,
$r_{\rm A}$ is the accretion radius of the hole and $\alpha$ is a dimensionless constant which depends on the equation of state ($p=k\rho$) and other features of the accretion flow. None of these arguments is precisely correct, since they all involve approximations which make them in some respects Newtonian. In particular, the value of $\alpha$ depends upon whether one allows for such effects as the relativistic sound speed, relativistic pressure and relativistic focussing. Also a stationary flow is usually assumed and the background cosmological expansion is neglected. Nevertheless it might be hoped that these inaccuracies could be absorbed into the constant $\alpha$.

If one assumes that the appropriate density to use in Eq.~(\ref{eq:newt_accretion}) is the background cosmological density $\sim 1/(Gt^2)$, the accretion rate becomes 
\begin{equation}
\frac{dm_{\rm BH}}{dt}=\frac{ m_{\rm BH}^{2}}{\beta t^{2}},
\label{eq:newtonian_accretion}
\end{equation}
where $\beta$ is another constant with units $c^{3}/G$. This has the following solution:
\begin{equation}
m_{\rm BH}=\frac{\beta t}
{1+\frac{t}{t_f} \left( \frac{\beta t_f}{M_f} -1\right)}
\label{eq:accretesoln}
\end{equation}
where $M_f$ is the mass of the black hole when it forms at time $t_f$. 
If the hole is much smaller than the particle horizon at formation ($M_f \ll \beta t_f$), this implies that there is very little growth. However,  if $M_f$ is chosen so that the term in brackets in Eq.~(\ref{eq:accretesoln}) is zero, it suggests the possibility of self-similar growth with $m_{\rm BH} \propto t$. 
We assume the constant $\beta$ is $6$, since that is what the analysis of
reference~\cite{hc2004a} implies when $k=1$.  If the Newtonian formula
were correct, the ratio would be order of unity.
See~\cite{hc2004a} for a full discussion of the solutions. 

In discussing the accretion rate for a scalar field, which is relevant here, one possibility is to regard it as a stiff fluid. In this case, one can apply the above analysis with $\beta$ having the value appropriate for $k=1$ and this is the approach used by Carr and Goymer~\cite{cg1999}. Bean and Magueijo~\cite{bm2003} also assume the accretion is given by Eq.~(\ref{eq:newt_accretion}) but identify $\rho$ with the kinetic energy of the scalar field ~$(1/2)\dot{\Psi}^2$. However, since this is assumed to falls off as $1/t^2$, they obtain an equation of the same form as Carr and Goymer, though with a different constant since they allow for a scalar potential.  Custodio and Horvath~\cite{ch2005} allow for a broader range of behaviours for $\dot{\Psi}$, leading to a generalization of Eq.(~\ref{eq:accretesoln}). All of these analyses suggest that the black hole could grow self-similarly with a fine-tuning of the initial mass. However, as discussed in the Introduction, this conclusion conflicts with the analytic demonstration that there is no self-similar solution which contains
a black hole attached to an exact flat Friedmann exterior 
solution~\cite{ch1974,lcf1976,bh1978a,bh1978b}. 

The prediction of a large accretion rate also conflicts with the results of our simulations. 
Since the cosmological time is given by $t=(2/3)[(u+v)/2]^{3/2}$
for the flat Friedmann solution
(see Appendix~\ref{subsec:flat_friedmann}), 
the self-similar growth of a black hole would imply that its 
mass grew like $(u+v)^{3/2}$.
The black hole mass accretion rate 
$dm_{\rm BHEH}/dv$
should therefore be proportional to $(u+v)^{1/2}$.
However, the results of our simulations 
show that the accretion rate decreases
after reaching a maximum.
We conclude that there is no evidence 
for a trend towards self-similar evolution,
which contradicts the argument~\cite{bm2003} that a black hole can grow as fast as the Universe through accreting a quintessence field. It also provides a counterexample to the proposal
that a spherically symmetric spacetime always evolves towards 
self-similarity~\cite{cc2000}.

It order to determine
the accuracy of the Newtonian accretion formula more precisely, 
we now derive an exact relativistic formula for the accretion rate.
Because the event horizon is given by 
an outgoing null surface $u=\mbox{const}$, the time variation of the
black hole mass is governed by Eq.~(\ref{eq:m_v}). Also 
Eqs.~(\ref{eq:rho_H}) and (\ref{eq:J}) imply 
\begin{equation}
\rho_{\rm H} + J = \frac{q^2}{4\pi a^{2}} = \frac{(\Psi_{,v})^2}{a^{2}},
\end{equation}
where $\rho_{\rm H}$ and $J$ are the energy density and momentum 
density. Combining this with Eq.~(\ref{eq:m_v}) gives
\begin{equation}
m_{,v} = -8\pi r^2 r_{,u} (\rho_{\rm H} + J).
\label{eq:exactaccrete}
\end{equation}
If we take the time coordinate to be
\begin{equation}
t=h(u+v),
\end{equation}
where $h$ is an arbitrary function of $u+v$, then we have
\[
\left. \frac{dm}{dt}\right|_{u={\rm const}} =m_{,v}\left.\frac{dv}{dt}\right|_{u={\rm const}}=\frac{m_{,v}}{h'},
\quad
\left.\frac{dr}{dt}\right|_{v={\rm const}} =r_{,u}\left.\frac{du}{dt}\right|_{v={\rm const}}=\frac{r_{,u}}{h'}.
\]
Combining this with Eq.(~\ref{eq:exactaccrete}) gives an expression for $(dm/dt)_{u={\rm const}}$. In particular, since the black hole event horizon has $u=$const, the accretion rate can be expressed as
\begin{equation}
\frac{dm_{\rm BHEH}}{dt} =-8\pi r^{2} (\rho_{\rm H} + J)\left.\frac{dr}{dt}\right|_{v={\rm const}}.
\end{equation}
This accretion equation is exact and might be compared with the 
Newtonian prediction given by Eq.~(\ref{eq:newt_accretion}). 
We see that $\rho$ is replaced by $\rho_{\rm H} + J$, while $c$ is
replaced by the value of $dr/dt$ along the path with constant $v$. 

Eq.~(\ref{eq:exactaccrete}) also explains why the mass accretion rate starts low for models A and D. Since the black hole event horizon is inside but very close to the cosmological apparent horizon in these cases, $r_{,u}$ is negative but very close to zero. After a while, however, $r_{,u}$ falls well below zero and then the mass accretion increases. The suppression of the 
accretion by the factor $r_{,u}$ 
on the right-hand side of Eq.~(\ref{eq:exactaccrete}) can be understood in a more general context
from the counterparts of Eqs.~(\ref{eq:m_u}) and (\ref{eq:m_v}) for a general
spherically symmetric spacetime~\cite{hc2004a,hayward1996}. Using the present notation, the equations become
\begin{eqnarray}
m_{,u}&=&\frac{8 \pi r^{2}}{a^{2}}(T_{uv}r_{,u}-T_{uu}r_{,v}), \\
\label{eq:m_u_general}
m_{,v}&=&\frac{8 \pi r^{2}}{a^{2}}(T_{uv}r_{,v}-T_{vv}r_{,u}).
\label{eq:m_v_general}
\end{eqnarray}
The combination of the two terms in parentheses on the right-hand side 
of Eq.~(\ref{eq:m_v_general}), related to outgoing and 
ingoing null expansions respectively,
determines the time variation of the black hole mass.
In the case of a massless scalar field, the situation is simplified 
because
$T_{uv}=0$ , $T_{uu}=(\Psi_{,u})^{2}\geq 0$ and 
$T_{vv}=(\Psi_{,v})^{2}\geq 0$.
As a result, we can immediately conclude that 
the black hole accretion vanishes when its event horizon coincides with
the cosmological apparent horizon, since
$r_{,u}=0$ there.

It can be easily proved that the black hole mass 
is non-decreasing if $r_{,v}>0$, $r_{,u}<0$ and 
if the dominant energy condition holds
on the event horizon (see Proposition 5 of \cite{hayward1996}).
However, if these assumptions are not satisfied,
the mass variation of the black hole event horizon is non-trivial. In particular, one finds that the 
black hole mass is decreasing when it is outside the 
cosmological apparent horizon, i.e., $r_{,u}>0$.
This possibility is studied in detail elsewhere~\cite{hc2004b}.
It should be noted that 
Hayward~\cite{hayward1996,hayward1993,hayward1994,hayward1998} introduced 
the future (past) outer trapping horizon
as a more general and useful concept than the black hole 
(white hole) event horizon.  When the black hole (white hole) horizon is defined in this way,
he also proved the monotonicity of the area and quasi-local mass in the 
spherically symmetric situation.

\subsection{Gravitational memory?}
The Brans-Dicke theory with empty stress-energy tensor
can be transformed into the Einstein theory with 
a massless scalar field by a conformal transformation
even when the Brans-Dicke parameter $\omega$ is of order unity~\cite{de1992}.
This is also the case in scalar-tensor theories of gravity
if there is a single massless gravitational scalar field.
The transformed frame with the Einstein-Hilbert action 
is called the Einstein frame, in contrast to the original 
physical frame. The gravitational constant varies in space and time
in the physical frame and is 
actually a function of the massless scalar field $\Psi$ 
in the Einstein frame.
Since the conformal transformation 
does not affect the causal structure of the spacetime,
the present simulation can be regarded 
as probing the evolution of PBHs 
in Brans-Dicke and scalar-tensor theories
with empty stress-energy tensor.

Our results elucidate whether or not gravitational memory 
is physically reasonable~\cite{cg1999}.
We have seen that the scalar field evolution 
near and even inside the black hole event horizon 
follows the cosmological evolution. Moreover, the gradient of the scalar field 
is always timelike. 
Therefore, we conclude that the evolution of 
the physical gravitational constant
near a PBH in Brans-Dicke and scalar-tensor theories
essentially follows its asymptotic cosmological evolution. 
The fact that the gradient of the scalar field 
is timelike during the evolution also ensures the existence
of a spacetime foliation with spacelike hypersurfaces,
in which the physical gravitational constant is spatially constant.
Therefore, we conclude that the gravitational memory scenario
is physically unrealistic even for PBHs
whose size is comparable with the cosmological horizon.

Using a perturbative test field analysis, Jacobson~\cite{jacobson1999} showed 
that black holes do not exhibit gravitational memory 
in scalar-tensor cosmology if they are much smaller than 
the cosmological horizon. Numerical studies
for a test scalar field 
in the PBH dust solution by Harada, Goymer and Carr~\cite{hgc2002}
showed 
that there is no gravitational memory even for black holes
whose size is comparable to the cosmological horizon scale, so long as the back reaction of the scalar field on the metric can be neglected. 
In the present work we have extended this result by considering the case where 
the black hole size is comparable to the cosmological horizon 
scale and the gravitational effect of the scalar field is included.
Once the black hole horizon gets
much smaller than the cosmological horizon
scale, the approximation made by Jacobson will apply.
Taken together, these results suggest that the
gravitational memory scenario is a highly unphysical, at least in Brans-Dicke and scalar-tensor theories.

\section{Conclusion}
We have investigated the evolution of a PBH
in a flat Friedmann universe with a massless scalar field
by numerically integrating the Einstein field equations
using the double-null formulation.
We have considered models in which a Schwarzschild interior
and a Friedmann exterior are initially matched at a finite radius with some 
smoothing length.
For PBHs which are initially 
the same size as or smaller than the cosmological apparent horizon,
the black hole event horizon soon becomes smaller than
the cosmological horizon scale.
The black hole apparent horizon forms inside the black hole event horizon
but asymptotes towards it.

The scalar field evolution qualitatively follows the
cosmological evolution close to the black hole event horizon
and even inside it.
The field maintains a timelike gradient and is therefore 
always equivalent
to a stiff fluid. In terms of the stiff fluid description,
the energy density in the perturbed region decreases
homogeneously in time and the infall velocity profile
near the black hole becomes almost stationary after 
a short while.  

Our results show that, soon after the PBH 
enters the cosmological apparent horizon,
it becomes isolated from the cosmological expansion,
with only a very small amount of mass accretion
from the non-vacuum exterior.
The accretion can be significant at first 
but soon decreases and becomes insignificant.
In particular, our simulations exhibit no self-similar growth
of the PBH.
This is consistent with earlier work, which proved the 
non-existence of self-similar PBH 
solutions for a stiff fluid 
with an exact flat Friedmann exterior.
The present result also indicates that the gravitational memory
scenario for PBHs is unphysical in 
Brans-Dicke and scalar-tensor theories of gravity. 

\acknowledgments
We would like to thank S.~A.~Hayward, J. Magueijo, J.~Miller 
and I.~Musco for helpful discussions.
TH was supported from JSPS.
\appendix
\section{physical quantities and numerical code}
\label{sec:basic_equations}

Following Hamad\'e and Stewart~\cite{hs1996}, we use variables,
$s$, $A$, $c$, $d$, $f$, $g$, $p$, $q$ as well as
auxiliary variables $\lambda$ and $\mu$ for numerical calculation, 
where $s=\sqrt{4\pi}\Psi$, $p=\sqrt{4\pi}\Psi_{,u}$,
$q=\sqrt{4\pi}\Psi_{,v}$, and $A=a^{2}$.
Their paper gives definitions and the full set of field equations 
for these quantities.
The physical quantities used in our discussion can be expressed
in terms of these quantities as follows:
\begin{eqnarray}
m&=&\frac{r}{2}\left(1+\frac{4fg}{a^{2}}\right), \\
\rho_{\rm H}&=&\frac{p^{2}+q^{2}}{8\pi a^{2}}, 
\label{eq:rho_H}\\
J&=&\frac{q^{2}-p^{2}}{8\pi a^{2}}, 
\label{eq:J}\\
\rho_{\rm S}&=&\frac{pq}{4\pi a^{2}}, \\
V&=&\frac{fq+gp}{a\sqrt{2a^{2}pq}}.
\end{eqnarray}

We use a two-step finite difference scheme slightly different from
the one proposed by Hamad\'{e} and Stewart~\cite{hs1996}. First we write the equations 
schematically as
\begin{eqnarray}
y_{,u}&=&F(y,z), \\
z_{,v}&=&G(y,z),
\end{eqnarray}
where 
\[
y=\left(\begin{array}{c}
q \\
d 
\end{array}
\right),
\quad
z=\left(
\begin{array}{c}
A\\
r\\
s\\
g\\
f\\
p
\end{array}
\right).
\]
At the first step we predict the values at $n(u,v)$
using those at $w(u,v-h)$ and $e(u-h,v)$ as
\begin{eqnarray}
\hat{y}_{n}&=&y_{e}+hF(y_{e},z_{e}), \\
\hat{z}_{n}&=&z_{w}+\frac{1}{2}h(G(y_{w},z_{w})+G(\hat{y}_{n},\hat{z}_{n})).
\label{eq:2nd_equation_1st_step}
\end{eqnarray}
At the second step we can correct the prediction via
\begin{eqnarray}
y_{n}&=&\frac{1}{2}(\hat{y}_{n}+y_{e}+hF(\hat{y}_{n},\hat{z}_{n})), \\
z_{n}&=&\frac{1}{2}(\hat{z}_{n}+z_{w}+hG(\hat{y}_{n},\hat{z}_{n})).
\label{eq:2nd_equation_2nd_step}
\end{eqnarray}
This is the scheme adopted by~\cite{hs1996}.
Due to Eq.~(\ref{eq:2nd_equation_2nd_step}), however, 
the scheme is only accurate to first-order, so we use instead 
\begin{equation}
z_{n}=\hat{z}_{n},
\end{equation}
since this is accurate to second-order and
does not suffer from numerical instabilities.

In general, 
Eq.~(\ref{eq:2nd_equation_1st_step}) 
can be implemented only implicitly.
However, in this special case, it is possible 
to determine $\hat{z}_{n}$ explicitly as follows:
\begin{eqnarray}
\hat{s}_{n}&=&\tilde{s}_{w}+\frac{1}{2}h\hat{q}_{n}, \\
\hat{A}_{n}&=&\frac{\tilde{A}_{w}}{1-
h\hat{d}_{n}}, \\
\left(\begin{array}{c}
\hat{r}_{n}\\
\hat{g}_{n}
\end{array}
\right)&=&
\frac{1}{1-h\hat{d}_{n}+\displaystyle{\frac{1}{4}}
h^{2}\hat{q}_{n}^{2}}
\left(\begin{array}{cc}
1-h\hat{d}_{n}&\displaystyle{\frac{1}{2}}h\\
-\displaystyle{\frac{1}{2}}h\hat{q}_{n}^{2}&1
\end{array}
\right)
\left(\begin{array}{c}
\tilde{r}_{w}\\
\tilde{g}_{w}
\end{array}
\right), \\
\hat{f}_{n}&=&\frac{\tilde{f}_{w}-\displaystyle{\frac{1}{8}}
h\frac{\hat{A}_{n}}
{\hat{r}_{n}}}
{1+\displaystyle{\frac{1}{2}}h\displaystyle{\frac{\hat{g}_{n}}
{\hat{r}_{n}}}}, \\
\hat{p}_{n}&=&\frac{\tilde{p}_{w}-\displaystyle{\frac{1}{2}}h
\displaystyle{\frac{\hat{f}_{n}\hat{q}_{n}}
{\hat{r}_{n}}}}
{1+\displaystyle{\frac{1}{2}}h\displaystyle{
\frac{\hat{g}_{n}}{\hat{r}_{n}}}},
\end{eqnarray}
where we put
\begin{equation}
\tilde{z}_{w}\equiv z_{w}+\frac{1}{2}hG(y_{w},z_{w}).
\end{equation}
Note that the order of determination is very important here.

\section{exact solutions in the double-null formulation}
\label{sec:exact_solutions}
\subsection{Flat Friedmann solution}
\label{subsec:flat_friedmann}
The flat Friedmann solution with a massless scalar field can be written as
\begin{equation}
ds^{2}=a^{2}[-d\eta^{2}+d\chi^{2}+\chi^{2}(d\theta^{2}+\sin^{2}\theta d\phi^{2})],
\end{equation}
where $a^{2}=C^{2}\eta$ and $C$ is a positive constant.
This solution can be rewritten in double-null coordinates as 
\begin{eqnarray}
ds^{2}&=&-a^{2}dudv+r^{2}(d\theta^{2}+\sin^{2}\theta d\phi^{2}), 
\label{eq:Friedmann1} \\
a^{2}&=&C^{2}\left(\frac{u+v}{2}\right), 
\label{eq:Friedmann2} \\
r&=&C\left(\frac{u+v}{2}\right)^{1/2}\left(\frac{v-u}{2}\right), 
\label{eq:Friedmann3} \\
s&=&\frac{\sqrt{3}}{2}\ln\left(\frac{u+v}{2}\right)+s_{0},
\label{eq:Friedmann4}
\end{eqnarray}
where $s_{0}$ is an arbitrary constant, and
$u$ and $v$ are related to 
$\eta$ and $\chi$ through
\begin{equation}
u=\eta-\chi, \quad v=\eta+\chi,
\end{equation}
respectively.
The cosmological time $t$, where $dt=a d\eta$,
$t$ is given by 
\begin{equation}
t=\frac{2}{3}C\eta^{3/2}=\frac{1}{3\sqrt{2}}C(u+v)^{3/2}.
\end{equation}
The Hubble parameter $H=d (\ln a)/dt$ is given by
\begin{equation}
H=\frac{1}{2C}\left(\frac{u+v}{2}\right)^{-3/2}.
\label{eq:hubble}
\end{equation} 

There are two important horizons in this spacetime:
the cosmological particle horizon and the cosmological
apparent horizon.
In the above coordinates, the cosmological particle horizon is given by 
$u=0$ or $\eta=\chi$, while the cosmological apparent horizon
is given by $3u+v=0$ or $\eta=\chi/2$.
Hence, the cosmological apparent horizon is 
spacelike and outside the cosmological particle horizon.
The conformal diagram of the flat Friedmann solution
is described in Fig.~\ref{fig:flat_friedmann}.

The physical quantities are
\begin{eqnarray}
m&=&\frac{C}{16\sqrt{2}}\frac{(v-u)^{3}}{(v+u)^{3/2}}, \\
\rho_{\rm H}&=&\frac{3}{8\pi C^{2}}\frac{1}{(u+v)^{3}}, \\
J&=&0, \\
\rho_{\rm S}&=&\frac{3}{8\pi C^{2}}\frac{1}{(u+v)^{3}}, \\
V&=&\frac{1}{2\sqrt{2}}\frac{v-u}{v+u}.
\end{eqnarray}

This solution admits the following
initial data:
\begin{eqnarray}
s(u_{0},v)&=&\frac{\sqrt{3}}{2}\ln\left(\frac{u_{0}+v}{2}\right)+s_{0}, \\
s(u,v_{0})&=&\frac{\sqrt{3}}{2}\ln\left(\frac{u+v_{0}}{2}\right)+s_{0}, \\
a^{2}(u_{0},v)&=&C^{2}\left(\frac{u_{0}+v}{2}\right), \\ 
a^{2}(u,v_{0})&=&C^{2}\left(\frac{u+v_{0}}{2}\right), 
\end{eqnarray}
and
\begin{eqnarray}
r(u_{0},v_{0})&=&C\left(\frac{u_{0}+v_{0}}{2}\right)^{1/2}
\left(\frac{v_{0}-u_{0}}{2}\right), \\
f(u_{0},v_{0})&=&r(u_{0},v_{0})\left[\frac{1}{2(u_{0}+v_{0})}
-\frac{1}{v_{0}-u_{0}}\right], \\
g(u_{0},v_{0})&=&r(u_{0},v_{0})\left[\frac{1}{2(u_{0}+v_{0})}
+\frac{1}{v_{0}-u_{0}}\right].
\end{eqnarray}

\subsection{Schwarzschild solution}
\label{subsec:schwarzschild}
The Schwarzschild solution is given by 
\begin{equation}
ds^{2}=-\left(1-\frac{2M}{r}\right)dt^{2}
+\left(1-\frac{2M}{r}\right)^{-1}dr^{2}+r^{2}(d\theta^{2}+\sin^{2}
\theta d\phi^{2}).
\end{equation}
This can be rewritten 
in double-null coordinates as
\begin{equation}
ds^{2}=-\left(1-\frac{2M}{r}\right)dudv+r^{2}(d\theta^{2}+\sin^{2}
\theta d\phi^{2}),
\end{equation}
where $u$ and $v$ are given by
\begin{equation}
u=t-r_{*}, \quad v=t+r_{*}, \quad \frac{dr_{*}}{dr}
=\left(1-\frac{2M}{r}\right)^{-1}.
\end{equation}
However, as is well known, 
this coordinate system has a coordinate singularity 
on the black hole event horizon.
We therefore consider another coordinate system.
For example, we can prescribe the initial data for this 
solution as follows:
\begin{eqnarray}
a^{2}(u_{0},v)&=&a^{2}(u,v_{0})=\mbox{const}, \\
s(u_{0},v)&=&s(u,v_{0})=s_{0}, \\
r(u_{0},v_{0})&=&r_{0}, \\
f(u_{0},v_{0})&=&-\frac{1}{2}, \\
g(u_{0},v_{0})&=&\frac{a^{2}}{2}\left(1-\frac{2M}{r_{0}}\right).
\end{eqnarray}

\newpage
\begin{table}[htbp]
 \begin{center}
\caption{\label{tb:models} Model parameters and initial mass
 within black hole event horizon compared to cosmological apparent horizon}
  \begin{tabular}{|c||c|c|c|c|c|c|c|}
   \hline
   Models & $u_{\rm m}$ & $\Delta u$ & $u_{0}$ & $u_{1}$ & $v_{0}$ & $v_{1}$ 
   & $m_{\rm BHEH}/m_{\rm CAH}$ \\ 
   \hline 
   A & $-1/3$ & 0.02 & $-0.5$ & 1.1 & 1 & 4 & 0.972\\
   B & 0     & 0.02 & $-0.5$ & 1.1 & 1 & 2 & 0.223\\
   C & $-1/3$ & 0.5 & $-0.5$ & 1.1 & 1 & 2.5& 0.399\\
   D & $-0.5$ & 0.5 & $-0.5$ & 1.1 & 1 & 3.5& 0.727\\
\hline
  \end{tabular}
 \end{center}
\end{table}
\begin{table}[htbp]
 \begin{center}
  \caption{\label{tb:radii} Initial and final radii of black hole horizons and their ratios}
  \begin{tabular}{|c||c|c|c||c|c|c|}
   \hline
   Models & BHAH $v=v_{0}$ & BHAH
   $v=v_{1}$ & Ratio & BHEH $v=v_{0}$ & BHEH $v=v_{1}$ & Ratio \\ 
   \hline 
   A & 0.373 & 0.668 & 1.79 & 0.377 & 0.694 & 1.84 \\
   B & 0.0858 & 0.124 & 1.44 & 0.0985 & 0.125 & 1.27 \\
   C & 0.154 & 0.232 & 1.51 & 0.173 & 0.241 & 1.39 \\
   D & 0.279 & 0.470 & 1.68 & 0.300 & 0.485 & 1.62 \\
\hline
  \end{tabular}
 \end{center}
\end{table}
\begin{table}[htbp]
 \begin{center}
  \caption{\label{tb:masses} Initial and final masses within black hole
  horizons and their ratios}
  \begin{tabular}{|c||c|c|c||c|c|c|}
   \hline
   Models & BHAH $v=v_{0}$ & BHAH $v=v_{1}$ & Ratio & BHEH $v=v_{0}$ & BHEH $v=v_{1}$ & Ratio \\ 
   \hline 
   A & 0.187 & 0.334 & 1.79 & 0.187 & 0.334 & 1.79 \\
   B & 0.0429 & 0.0614 & 1.43 & 0.0429 & 0.0614 & 1.43 \\
   C & 0.0769 & 0.116 & 1.51 & 0.0769 & 0.116 & 1.51\\
   D & 0.140 & 0.235 & 1.68 & 0.140 & 0.235 & 1.68 \\
\hline
  \end{tabular}
 \end{center}
\end{table}



\end{document}